\documentclass{aa}
\usepackage{psfig}
\begin{document}

\title{A Spectro-astrometric study of southern pre-main sequence stars}

\subtitle{ --- binaries, outflows, and disc structure down to
           AU scales ---}

\author{Michihiro Takami\inst{1}     \and
        Jeremy Bailey\inst{2}        \and
        Antonio Chrysostomou\inst{1}}

\institute{Department of Physical Sciences, University of
           Hertfordshire, Hatfield, Herts AL10 9AB, UK
           \and
           Anglo-Australian Observatory, PO Box 296, Epping, NSW 1710,
           Australia}

\offprints{Michihiro Takami, \email{takami@star.herts.ac.uk}}

\date{Received 00 ?? 2002 / Accepted ?? ?? 2002}

\abstract{
We present spectro-astrometric observations for 28 southern pre-main sequence (PMS) stars and investigate their circumstellar environment down to AU scales. The structures detected in the ``position spectra'' include: (1) almost all the known binary companions in our sample (Sz 68, Sz 41, HO Lup, VW Cha, S CrA, AS 205), (2) companion candidates which have not been detected by infrared speckle techniques (T CrA, MWC 300), (3) monopolar and bipolar jets (AS 353A, CS Cha), (4) a combination of jets and a bow shock (VV CrA), and (5) a combination of a jet and stellar companion (R CrA). Results in known binaries show that this technique is capable of detecting binaries with separations down to $\sim$10 milliarcsec.
Both components in each binary appear to have strikingly similar profiles in H$\alpha$ emission, indicating a similarity of circumstellar activity (mass accretion and/or a wind), and supporting the scenario of core fragmentation for the mechanism of binary formation. The bipolar H$\alpha$ jet in CS Cha has a spatial scale of $\sim$1.5 AU, similar to that previously observed in RU Lup, and likely be heated by a mechanism other than shocks. From the spatial scale, velocity, and H$\alpha$ luminosity, we estimate the mean hydrogen density in the AU-scale bipolar flows to be $\ga$10$^7$ cm$^{-3}$. The bipolar geometry in these jets can be explained by the presence of a disc gap/hole at AU scales, which could be induced by a gas-giant planet at the ice condensation radius.

\keywords{
accretion, accretion discs --
line: formation
line: profiles
stars: activity
stars: evolution
stars: pre-main-sequence
}
}

\titlerunning{Southern pre-main sequence stars}

\authorrunning{Takami et al.}

\maketitle

\section{Introduction}
The circumstellar environment within 10 AU of young stars are of particular interest for star and planet formation. In this region, the star accretes the material from the circumstellar disc, and drives an energetic jet/wind (see, e.g., Najita et al. 2000; Shu et al. 2000, K\"onigl \& Pudritz 2000). In the circumstellar disc, planets would be growing up and tidally interact with the circumstellar environment (see e.g., Lubow \& Artimowicz 2000). This region also includes a stellar companion in a significant number of objects, and the population, mass distribution, and orbital parameters of these companions could give useful constraints on the understanding of binary formation (see e.g., Mathieu et al. 2000). However, even the $Hubble$ $Space$ $Telescope$ or adaptive optics on 10-m telescopes cannot resolve this region: a spatial scale of 10 AU corresponds to 0".07 in the nearest star forming regions, which is comparable to the angular resolutions of these facilities. Observational studies in this region have thus been carried out by photometry and spectroscopy without any direct spatial information, and optical-to-radio interferometry in a limited number of objects.

To obtain spatial information well below these resolutions, we have begun ``spectro-astrometric'' observations of pre-main sequence (PMS) stars. Bailey (1998a) applied this technique for a PMS star, evolved star, and AGN, showing that the technique allows us to study the spatial structure of their emission/absorption features down to $milliarcsec$ scales. Bailey (1998b) succeeded to detect almost all known PMS binaries in his sample, proving the potential of this technique to find new binaries with separations of well below 0.1 arcsec. Takami et al. (2001) then revealed the presence of bipolar structure at $\sim$30 mas scale in the outflow of the T Tauri star RU Lup. This bipolar structure contrasts with the jet at larger scales, which exhibits only blueshifted components,
suggestive of the presence of a disc gap at a few AU which allows the redshifted H$\alpha$ outflow to be seen.

In this paper, we present results of spectro-astrometric observations for 28 southern PMS stars, and investigate their binarity, outflow activity, and disc structure down to AU scales. Section 2 summarizes the observations. Sections 3 shows the detected structures due to stellar companions, jets, and bow shocks. Section 4 shows the H$\alpha$ profiles for each component of known binaries, which probes the similarity or difference of circumstellar activities between binary components.
Section 5 investigates the detectability of PMS binaries, the mechanism of binary formation, physical conditions of the H$\alpha$ outflows down to AU scales, and the possibility of planet formation in a few PMS stars. The conclusions of this paper is described in Section 6.

\section{Observations}
Observations were carried out
at the 3.9-m Anglo-Australian Telescope using the RGO spectrograph.
The configuration with a 1-arcsec slit width, a 1200 line mm$^{-1}$
grating, and the 82-cm camera provided a spectral resolution
($\lambda / \Delta \lambda$) of 7000.
The pixel scale of 0.15 arcsec with the MITLL CCD
(2048$\times$4096 deep depletion CCD from MIT Lincoln Labs)
provides good sampling of the seeing profile, whose FWHM varies between 1 and 3 arcsec during the observations.
The flat fields were made by combining
many exposures with the spectrograph illuminated by a tungsten lamp.
Wavelength calibration were made by observations of a CuAr lamp.

Table 1 shows the log of the observations.
The targets were selected based on their brightness ($V \la 14$),
and most of them lie in the nearest low-mass star forming regions
including Lupus (190 pc --- Wichmann et al. 1998), Chamaeleon (170 pc --- Bertout et al. 1999), Corona-Australis (130 pc --- Marraco \& Rydgren 1981), Ophiuchus (140 pc --- de Geus \& Burton 1991), and Upper Scorpius (150 pc --- de Zeeuw et al. 1999).
The spectra were obtained with a wavelength coverage of
6480--6780 {\AA} (but 6140--6800 {\AA} for VV CrA)
to observe lines including H$\alpha$, \hbox{[S\,{\sc ii}]},
\hbox{He\,{\sc i}} 6678 {\AA}, and obtained at four slit position
angles (0$^{\circ}$, 90$^{\circ}$, 180$^{\circ}$, and 270$^{\circ}$)
to determine the positional displacement in N-S and E-W directions.
Although several targets were previously observed by Bailey (1998b),
we observed them again to ensure all the data had uniform quality with which to carry out a statistical study.

The data were reduced using the FIGARO package provided by Starlink.
After subtracting the bias level and flat-fielding,
the position spectrum was determined by fitting the seeing profile
at each wavelength with a Gaussian function.
After the fitting, the position spectra with opposite
position angles (0$^{\circ}$--$180^{\circ}$, or 90$^{\circ}$--$270^{\circ}$) were subtracted to remove instrumental effects
(see Bailey 1998a,b and Takami et al. 2001 for detail).
Each position spectrum has an arbitrary zero point, which
we adjust to correspond to the continuum position.
In addition to the position spectra, the intensity spectra were
obtained by subtracting the bias, flat-fielding, subtracting
the adjacent sky and extracting bright columns on the CCD. 
The systemic velocity of the targets was calibrated using photospheric absorption lines including \hbox{Li\,{\sc i}} 6707.82 {\AA}, \hbox{N\,{\sc ii}} 6613.62 {\AA}, \hbox{Ca\,{\sc i}} 6439.08 {\AA}, \hbox{Fe\,{\sc i}} 6475.63 {\AA}, and \hbox{Fe\,{\sc ii}} 6456.38 {\AA}.

The measured positional accuracy, $\sigma$,
depends on the seeing size and the photon number at each wavelength.
Each photon should be within a distance of the order of the seeing size
from the actual position of the source, and this positional uncertainty is reduced by a factor of $N^{1/2}$ by observing $N$ photons.
Consequently, the detection limit is described as follows
(Bailey 1998a):
\begin{equation}
\sigma{\rm{(milliarcsec)}} = const. \cdot \rm{FWHM(milliarcsec)} \cdot {\it N}^{-1/2},
\end{equation}
where FWHM is the full width half maximum of the seeing.
Table 1 provides the constant value of 0.5$\pm$0.2, in agreement
with that estimated using a Gaussian seeing profile (0.4).
These results imply that the positional accuracies in our spectra are dominated by photon noise, other noise (e.g., readout noise)
being negligible.

\begin{table*}
 \centering
 \begin{minipage}{140mm}
 \caption{Log of the observations}
 \begin{tabular}{@{}lcccclccc@{}}
Object		  & HBC%
  \footnote{the number in the Herbig\& Bell (1988) catalogue} & type%
  \footnote{references --- Herbig \& Bell (1988), Leinert et al. (1997)} & V & Distance%
  \footnote{references --- Bertout et al. (1999), de Geus \& Burton (1991), de Zeeuw et al. (1999), Herbig \& Jones (1983), Hillenbrand et al. (1992), Leinert et al. (1997), Marraco \& Rydgren (1981), Whichmann et al. (1998)
} & \multicolumn{1}{c}{Date}  & seeing & Photons%
\footnote{for each three pixels (corresponding to 0.4 {\AA}) along the wavelength.}
 & 1-$\sigma$ Detection Limit \\
                  &     &      &(mag)& (pc)&    & (") & ($\times 10^4$) & (mas) \\ [10pt]
SZ Cha		  & 566 &  K0  &12.0&~~170&1999.7.2& 2.8 & ~2   & 12.6 \\
Th 12    	  & 605 &  M0  &11.9&~~190&1999.7.2& 2.1 & ~6   & ~~2.9  \\
MWC 300		  & --- &  B   &10.5&1550&1999.7.2& $<$1.9\footnote{these objects show asymmetric profiles, suggesting that a stellar companion contributes to the measured FWHM} & ~4   & ~~3.6  \\
T CrA\footnote{this object was observe twice to investigate the time variation of the spatial structure reported by Bailey (1998a)}		  & 290 &  F0  &13.4&~~130&1999.7.2& $<$2.6$^e$ & ~4   & ~~4.6  \\
S CrA		  & 286 &  K6  &11.1&~~130&1999.7.2& $<$2.9$^e$ & ~2   & ~~6.7  \\
LkH$\alpha$ 332-20& 244 &  K2  &11.2&~~170&1999.7.3& 1.2 & ~9   & ~~1.3  \\
CS Cha		  & 569 &  K5  &11.6&~~170&1999.7.3& 1.2 & ~8   & ~~1.6  \\
Sz 77		  & 603 &  M0  &12.5&~~190&1999.7.3& 0.9 & ~8   & ~~1.5  \\
SR 20		  & 643 &$<$K7 &14.1&~~140&1999.7.3& 1.4 & ~2   & ~~3.9  \\
LkH$\alpha$ 118	  & 281 &  B5  &11.1&~~---&1999.7.3& 1.8 & ~6   & ~~3.3  \\
R CrA		  & 288 &  A5  &10.7&~~130&1999.7.3& 4.5 & ~2   & 15.3 \\
LkH$\alpha$ 332-21& 247 &  G8  &11.0&~~170&1999.7.4& 1.0 & 14  & ~~1.1  \\
Sz 41		  & 588 &  K0  &11.6&~~170&1999.7.4& 1.1 & ~7   & ~~1.9  \\
Sz 68		  & 248 &  K2  &10.3&~~190&1999.7.4& 1.2 & 23  & ~~0.9  \\
GQ Lup		  & 250 &  K7  &11.4&~~190&1999.7.4& 1.3 & ~7   & ~~2.1  \\
VV Ser		  & 282 & B-A  &11.9&~~440&1999.7.4& 1.2 & ~8   & ~~2.6  \\
VV CrA		  & 291 &  K7  &13.0&~~130&1999.7.4& 1.2 & ~1   & ~~5.0  \\
VW Cha            & 575 &  K2  &12.5&~~170&2000.7.21& 1.7 & ~1  & ~~8.2  \\
AS 205            & 254 &  K5  &12.4&~~150&2000.7.21& 2.0 & ~2  & ~~5.6  \\
TW Hya            & 568 &  K7  &10.9&~~~~56&2000.7.21& 1.3 & ~4  & ~~2.2  \\
AS 353A           & 292 &  --- &12.5&~~300&2000.7.21& 2.3 & ~1  & ~~7.7  \\
T CrA$^f$	  & 290 &  F0  &13.4&~~130&2000.7.21& $<$2.6$^e$ & ~2  & ~~4.6  \\
V4046 Sgr         & --- &  --- &10.5&~~---&2000.7.21& 2.8 & ~4 & ~~4.5 \\
CU Cha            & 246 &  A0  &~~8.5 &~~170&2000.7.21& 1.8 &1--2& ~~1.9 \\
AS 209            & 270 &  K5  &11.5&~~150&2000.7.21& 2.6 & ~3 & ~~4.6 \\
HO Lup            & 612 &  M1  &12.9&~~190&2000.7.23& 1.1 & 0.5& ~~6.7  \\
CT Cha            & 570 &  K7  &12.4&~~170&2000.7.23& 1.3 & ~2  & ~~2.5  \\
BF Cha            & --- & ---  & ---&~~170&2000.7.23& 1.3 & ~3 & ~~2.8 \\
AS 206            & 259 & K6-7 &12.9&~~140&2000.7.23& 1.4 & ~1 & ~~5.8 
\end{tabular}
\end{minipage}\
\end{table*}

\section{Positional displacement}
Displacement in the position spectra of 12 objects were detected, including: (1) known binaries with separations of 0.1--1.5 arcsec (Sz 68, Sz 41, HO Lup, VW Cha, S CrA, and AS 205); (2) T CrA and MWC 300, whose displacement can be explained by a stellar companion; (3) CS Cha, AS 353A, and VV CrA, whose displacement can be attributed to outflowing gas; and (4) R CrA, whose displacement can be attributed to a combination of a stellar companion and outflowing gas. The following subsections describe the details for individual cases.

\begin{figure*}
  \begin{center}
    \leavevmode
  \psfig{file=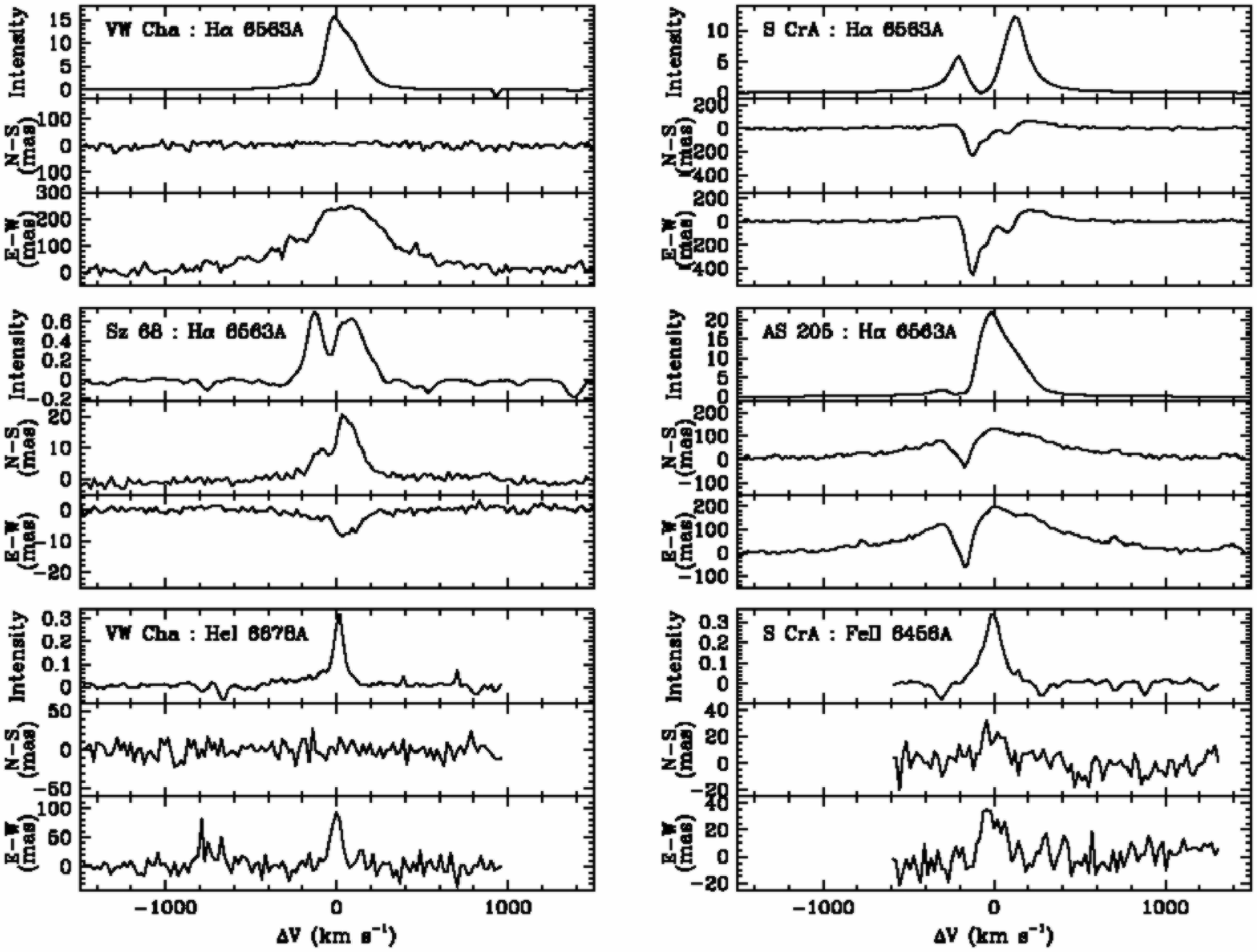, clip=,width=15 cm}
  \end{center}
\caption{Examples of intensity and position spectra for known binaries. Intensity spectra are normalized to the continuum.
}
  \label{knownbs1}
\end{figure*}

\begin{figure*}
  \begin{center}
    \leavevmode
\psfig{file=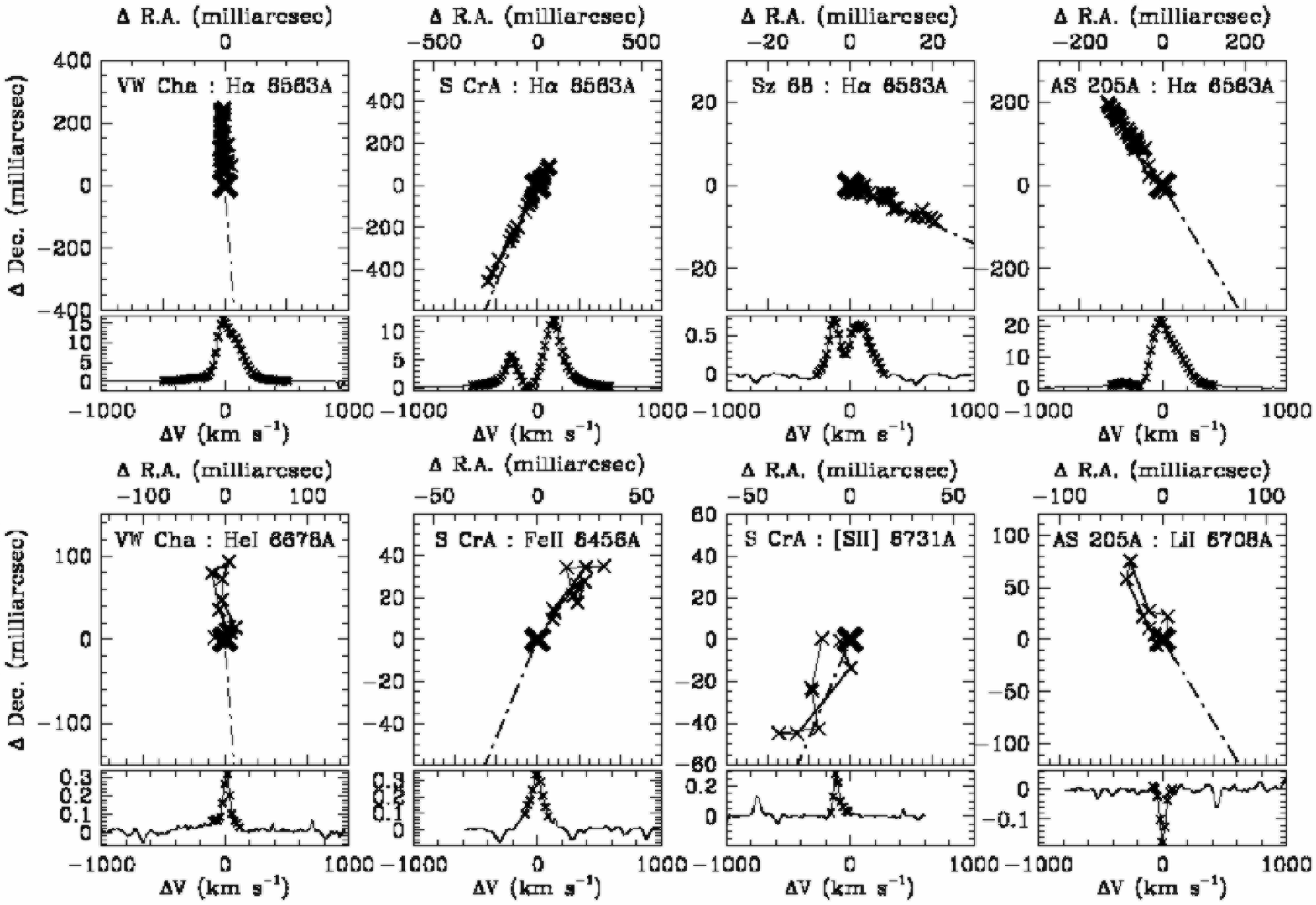,clip=,width=15 cm}
 \end{center}
\caption{Intensity profiles and positional displacement in emission/absorption features of known binaries VW Cha, S CrA, Sz 68, and AS 205. Small crosses in each map correspond to the velocity components shown in the intensity profile. A large cross in each figure indicates the centroidal position of the continuum source. A dot-dashed line indicates the direction of the fainter companion in the $K$-band continuum (Ghez et al. 1993,1997a). The intensity profiles are normalized to the continuum.
}
  \label{knownbs2}
\end{figure*}

\subsection{Known binaries (Figs \ref{knownbs1} and \ref{knownbs2})}

Among seven known binaries with separations of
0.1--1.5 arcsecs, displacement in H$\alpha$ emission
was detected in all of them but SR 20.
Some of them also show displacement
in other features including \hbox{Fe\,{\sc ii}} 6456 \AA, \hbox{Fe\,{\sc ii}} 6516 \AA,
\hbox{He\,{\sc i}} 6678 \AA, and \hbox{Li\,{\sc i}} 6708 \AA.

The position angles and the angular scales of the displacement are summarized in Table 2.
All the observed displacements lie at the binary position angle,
and this tendency can be attributed to the two binary components having different spectra (Bailey 1998b). The exception is H$\alpha$ emission in S CrA and AS 205, in which the direction of the displacement differs from the binary position angle beyond the 1-$\sigma$ level.
A likely explanation for S CrA
is that the two binary components were not equally within the slit
due to their wide binary separation. Indeed, infrared
speckle observations give a binary separation of
1.4 arcsecs in this object, which is comparable
to the seeing size for our observations (see Table 1).
One might think that the observed difference of the position angle
is due to orbital motion, although it is unlikely for the
following reasons: (1) the binary position angle in S CrA increases
with time (Bailey 1998b) while
our results show a smaller position angle than previous
measurements of the binary position angle by Ghez et al. (1997a),
and (2) a central mass of 30--40 M$_{\odot}$ would be required
to explain the difference between the two position angles,
not plausible as a mass for a PMS star ($\la$8 M$_{\odot}$ --- Stahler \& Walter 1993). On the other hand, the different position angles in AS 205 can be explained by the orbital motion of the system. Indeed, the difference between the two position angles suggest
a mass of the primary star of 2.4 M$_{\odot}$ assuming a circular orbit, consistent with
that derived from the stellar luminosity and spectral type
(1.5--2.5 M$_{\odot}$ --- Liu et al. 1996).

\begin{table*}
 \centering
 \begin{minipage}{160mm}
 \caption{Positional displacement in known binaries}
 \label{tab:table}
 \begin{tabular}{@{}lrrlcrcc@{}}
object & \multicolumn{1}{c}{separation\footnote{observed by Ghez et al. (1995,1997a)}}  & \multicolumn{1}{c}{P.A.$^a$} & \multicolumn{2}{c}{line} &  \multicolumn{2}{c}{observed angular scale}  & observed P.A.  \\ 
& \multicolumn{1}{c}{(mas)} & \multicolumn{1}{c}{(deg.)} & & & (milliarcsec) & (per binary separation) & (deg.) \\[10pt]
Sz 68   &$~107 \pm ~~7$&$245 \pm ~7$&H$\alpha$&6563 \AA&$ ~21 \pm  ~1$& 0.20&$248.2 \pm 0.9$\\
Sz 41   &$1500 \pm 800$&$150 \pm 20$&H$\alpha$&6563 \AA&$ ~22 \pm  ~3$& 0.01&$186   \pm 3$\\
HO Lup  &$1490 \pm ~70$&$ ~35 \pm ~2$&H$\alpha$&6563 \AA&$ ~24 \pm  ~9$& 0.02&$ 213   \pm 1$\\
VW Cha  &$~660 \pm ~30$&$184 \pm ~2$&H$\alpha$&6563 \AA&$250 \pm 12$& 0.38&$  2.5 \pm 0.5$\\
        &              &            &\hbox{He\,{\sc i}}      &6678 \AA&$ ~93 \pm 12$& 0.14&$  4   \pm 3$\\
S CrA   &$1410 \pm ~60$&$157 \pm ~2$&H$\alpha$&6563 \AA&$513 \pm ~9$& 0.36&$153.1 \pm 0.3$\\
        &              &            &         &        &$111 \pm ~9$& 0.08&$~331.1 \pm 0.5$\\
        &              &            &\hbox{Fe\,{\sc ii}}     &6456 \AA&$ ~47 \pm ~9$& 0.03&$323 \pm  3$\\
        &              &            &\hbox{Fe\,{\sc ii}}     &6516 \AA&$ ~49 \pm ~9$& 0.03&$338 \pm  4$\\
        &              &            &\hbox{He\,{\sc i}}      &6678 \AA&$ ~57 \pm ~9$& 0.04&$150 \pm  5$\\
        &              &            &\hbox{[S\,{\sc ii}]}    &6716 \AA&$ ~31 \pm ~9$& 0.02&$146 \pm  2$\\
        &              &            &\hbox{[S\,{\sc ii}]}    &6731 \AA&$ ~57 \pm ~9$& 0.04&$148 \pm  3$\\
AS 205 &$1317 \pm ~~2 $&$211.87 \pm 0.08$&H$\alpha$&6563 \AA&$239 \pm  ~8$& 0.18&$ 33.5 \pm 0.4$\\
        &               &                 &         &        &$~67 \pm  ~8$& 0.05&$ 202  \pm   4$\\
        &              &            &\hbox{Fe\,{\sc ii}}     &6456 \AA&$~61  \pm ~8$& 0.05&$ 36 \pm 3$\\
        &              &            &\hbox{Fe\,{\sc ii}}     &6516 \AA&$~60  \pm ~8$& 0.05&$ 37 \pm 2$\\
        &              &            &\hbox{He\,{\sc i}}      &6678 \AA&$~34  \pm ~8$& 0.03&$ 34 \pm 4$\\
        &              &            &\hbox{Li\,{\sc i}}      &6708 \AA&$~81  \pm ~8$& 0.06&$ 28 \pm 3$
\end{tabular}
\end{minipage}
\end{table*}

The displacement of H$\alpha$ emission/absorption in Sz 68, VW Cha, S CrA, and AS 205 show angular scales of 0.2--0.4 times as large as the binary separation, while it is much smaller in HO Lup and Sz 41 (0.01--0.02 as large as the binary separation). The latter results could be due to
their large binary separations ($\sim$1.5 arcsec), in which
the seeing of the brighter component does not include
the flux from the fainter component.
We thus conclude that angular scales for
H$\alpha$ emission in all the objects are consistent with those
obtained in other pre-main sequence stars by
Bailey (1998b): that is, 0.07--0.7.
The displacement in most of the objects
indicates the direction for one of the binary components,
while that in the H$\alpha$ profiles in S CrA and AS 205
lies in both directions: these profiles show
central absorption and wing emission features, in which
the position is displaced in the opposite directions.

The angular scales of the displacement in other emission/absorption lines
are 0.02--0.05 per binary separation, much smaller than that in H$\alpha$ emission.
These small scales are attributed to
smaller line-to-continuum ratios,
which allow larger contribution of the continuum flux and
reduce the positional displacement. In S CrA, the displacement indicates different directions between the lines: two \hbox{Fe\,{\sc ii}} lines lie in the direction of the primary, while \hbox{He\,{\sc i}} and \hbox{[S\,{\sc ii}]} lines lie in the direction of the secondary.

Positional displacement of forbidden lines in PMS stars is often attributed
to a jet/wind (e.g., Hirth, Mundt, \& Solf 1997; Takami et al. 2001,2002),
however, this explanation is not likely for the \hbox{[S\,{\sc ii}]} lines
in S CrA. 
Indeed, optical polarization of T Tauri stars
is very often perpendicular to jets within a few degrees
(M\'{e}nard \& Bastien 1992), although that in S CrA
was measured to be 151--156 degree (Bastien 1985),
nearly parallel to that of the positional displacement
in the \hbox{[S\,{\sc ii}]} emission. We thus conclude that
the displacement of the \hbox{[S\,{\sc ii}]} lines of S CrA is
due to a binary companion.



\begin{figure*}
  \begin{center}
    \leavevmode
\psfig{file=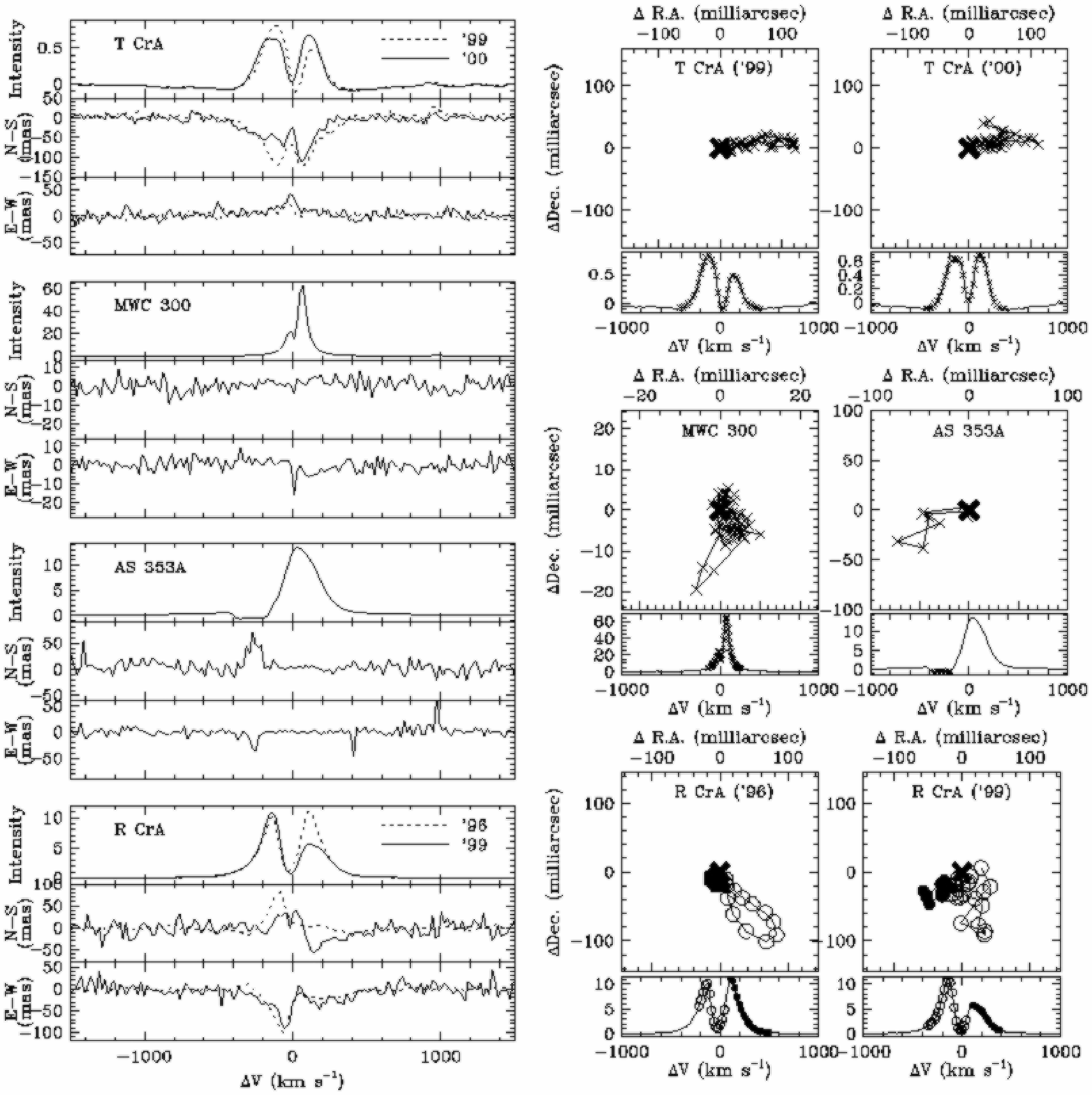,clip=,width=15 cm}
 \end{center}
\caption{Intensity profile and position displacement of the H$\alpha$ emission in T CrA, MWC 300, CS Cha, AS 353A, and R CrA. The intensity profiles are normalized to the continuum. The results of R CrA in 1996 was obtained by Bailey (1998b).
}
  \label{others}
\end{figure*}

\subsection{T CrA (Fig.~\ref{others})}

Infrared speckle observations by Ghez et al. (1997a) did not show the presence of a stellar companion, whereas the position spectra show displacement similar to that in known binaries, as first reported by Bailey (1998b). His and our results show 
angular scales for the displacement of 76$\pm$5, 140$\pm$9, 122$\pm$5, and
113$\pm$3 milliarcsec in 1996, 1997, 1999, and 2000, respectively,
indicating a lower limit for the binary separation of 140 milliarcsec.
Such a separation is larger than the detection limit of infrared speckle, thus the non-detection of the companion by Ghez et al. (1997a) implies that the companion has a $K$-magnitude fainter than 10.5.

\begin{figure}
  \begin{center}
    \leavevmode
\psfig{file=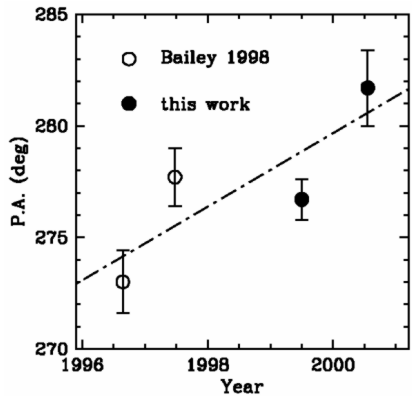,clip=,width=5.5 cm}
 \end{center}
\caption{Position angles of the displacement measured in T CrA. A dot-dashed line shows the regression line with an inclination of 1.65$^\circ$ yr$^{-1}$.
}
  \label{TCrA}
\end{figure}

Fig.~\ref{TCrA} shows the position angles of the displacement measured in our four
observing runs.
The angle increases with time, corresponding to $\sim$2 degree/yr,
consistent with the orbital motion of the stellar companion.
From this result, we estimate a binary separation of
$\sim$260 milliarcsec adopting a primary stellar mass of 0.8 M$_{\odot}$
(Hillenbrand et al. 1992) and assuming a circular orbit.
In this case, the angular scales observed in the four runs
would correspond to 0.29--0.54 per binary separation, consistent with
those observed in known binaries (\S 3.1).
On the other hand, time variation of the observed angular scale is not systematic as expected for binary motion, and such results could be attributed to variability of the emission feature
between binary components (Bailey 1998b). Indeed, H$\alpha$ emission observed by Bailey (1998b) and this work exhibit different shapes of profiles, supporting this explanation.

\subsection{MWC 300 (Fig.~\ref{others})}
Positional displacement was detected in the central absorption feature of the H$\alpha$ emission, with an angular scale and a position angle of 17$\pm$4 milliarcsec
and $\sim$165$^{\circ}$, respectively.
Similar positional displacement was also observed in the known binary S CrA; thus, it is likely that
the observed displacement in MWC 300
is due to a stellar companion. One might think that the observed displacement
could be explained by a jet/wind,
however, it is not likely since the
displacement in this object is observed only at positive velocities, different
from those expected in jets/winds (e.g., Takami et al. 2001,2002), where the displacement is observed at negative or both velocities.

Previous infrared speckle observations failed to detect this companion
(Leinert, Richichi, \& Haas 1997; Pirzkal, Spillar, \& Dyck 1997).
Corporon \& Lagrange (1999) argue that MWC 300 is a spectroscopic binary,
although our results cannot be explained as such:
a combination of a lower limit for the binary separation provided by
our results, the distance to the star (15.5 kpc --- Leinert et al. 1997), and
a typical mass of Herbig AeBe stars (5 M$_{\odot}$) suggests an orbital period of more than $2 \times 10 ^ 3$ yrs,
much larger than the duration of the observations by
Corporon \& Lagrange (1999).

\begin{figure*}
  \begin{center}
    \leavevmode
\psfig{file=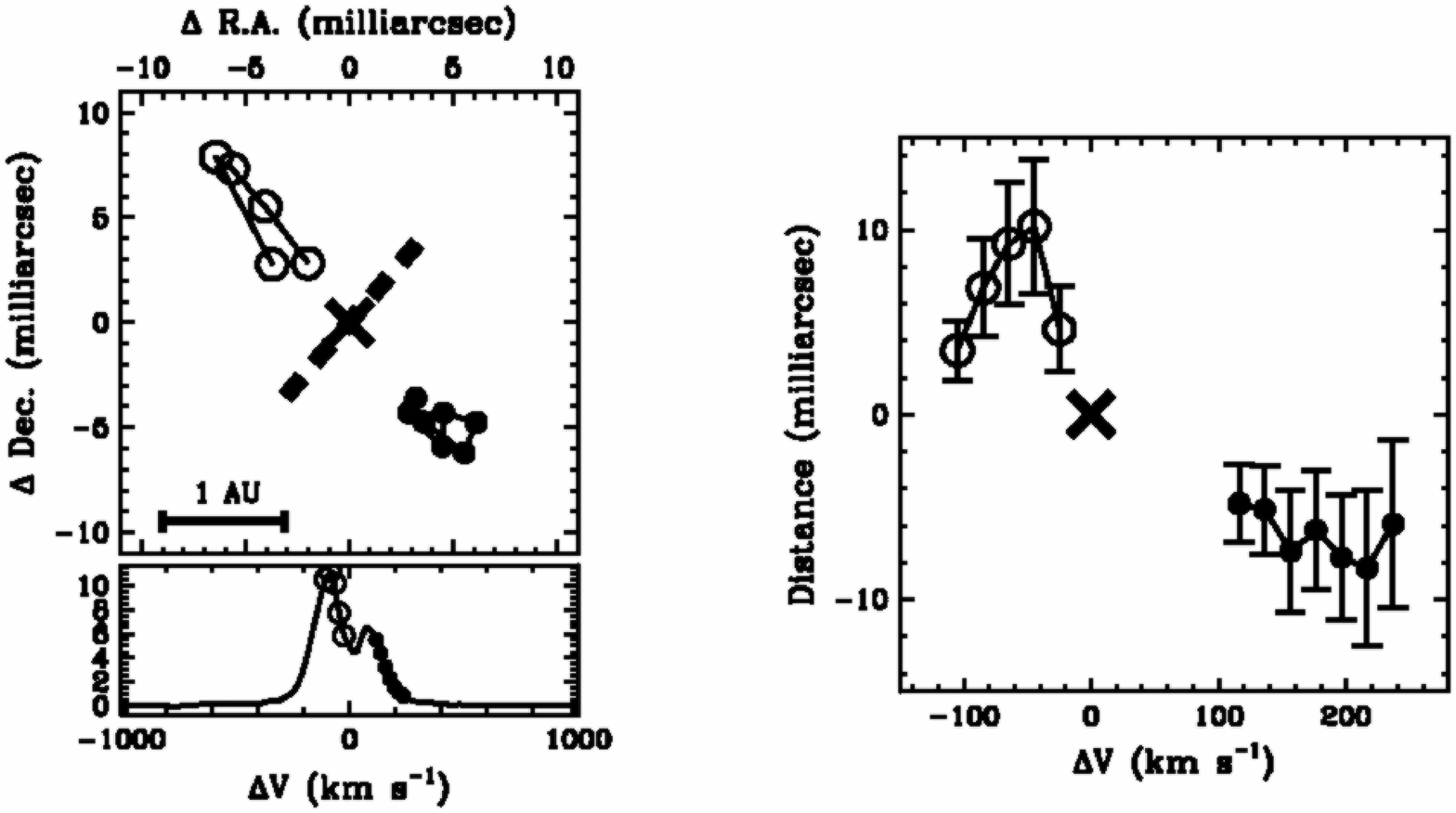,clip=,width=12 cm}
 \end{center}
\caption{(left) Spatial distribution of positional
displacement in H$\alpha$ emission of CS Cha. Contamination by the continuum
was removed using the line-to-continuum ratio at each wavelength: thus, the
plot shows the true angular scale of the H$\alpha$ emission. Open and filled
circles show blue- and redshifted components, respectively, and
each circle in the map corresponds to that in the intensity profile shown below.
A large cross and a dashed bar at the centre show the centroidal position
of the continuum and the direction of optical polarization,
respectively. The scale of the profile is normalized to the continuum
flux.
(right) Correlation between velocity and radial
distance of the displacement. A large cross
shows the centroidal position of the continuum.
}
  \label{CSCha}
\end{figure*}

\subsection{CS Cha (Figs \ref{others} and \ref{CSCha})}

The displacement in H$\alpha$ emission shows a bipolar structure: the blueshifted component
extends towards the north-east while the redshifted component extends towards
the south-west. The direction of the displacement is nearly
perpendicular to that of the optical polarization (139$^\circ$ ---
Drissen, Bastien, \& St.-Louis 1989), which is very often perpendicular to that
of jets (M\'{e}nard \& Bastien 1992, and references therein).
Thus, it is likely that the bipolar structure in CS Cha is due to
outflowing gas, as Takami et al. (2001) argue for RU Lup.

To determine the true spatial scale of the bipolar flow, we removed
contamination  by the continuum using the line-to-continuum ratio at each
wavelength (see Takami et al. 2001 for detail), and plot the results
in Fig.~\ref{CSCha}. The figure shows an angular scale of 8--10 milliarcsec,
and asymmetric  velocity field between the blueshifted
and redshifted component: the displacement is observed at
velocities of $-$100 to $-$20 km s$^{-1}$ in the blueshifted component
and $+100$ to $+250$ km s$^{-1}$ in the redshifted component. The displacement
along the outflow axis does not show clear positive correlation with
the velocity as observed in RU Lup by Takami et al. (2001).
Thus, the velocity field observed in this object cannot be simply attributed to
the acceleration of the outflow, in which the displacement should increase with
the velocity. Our result could be due to a contribution of magnetospheric mass accretion columns (see, e.g., Muzerolle, Calvet, \& Hartmann 1998,2001) or an inner hot wind (Beristain, Edwards, \& Kwan 2001; Alencar \& Batalha 2002) to the observed H$\alpha$ flux.

Forbidden lines in PMS stars often exhibit only blueshifted components due to obscuration of the counter flow by a circumstellar disc.
(Eisl\"{o}ffel et al. 2000, and references therein).
On the other hand, H$\alpha$ outflows in CS Cha have both blueshifted and
redshifted components. This result can be explained if a gap or hole in the circumstellar disc allows the redshifted components to be seen (Takami et al. 2001).
Indeed, the infrared spectral energy distribution (SED) in CS Cha
shows no substantial excess emission less than 20 $\mu$m
(Gauvin \& Strom 1992), indicating a
lack of dust with temperatures of more than 200 K.
If we adopt a stellar luminosity of 2.7 L$_{\odot}$
(Gauvin \& Strom 1992), this temperature corresponds to a disc radius
of 6 AU, consistent with the spatial scale of the observed redshifted H$\alpha$ outflow, i.e., $\sim$1.4 AU.
%

\subsection{AS 353A (Fig.~\ref{others})}
A positional displacement was observed within the blueshifted
absorption of the H$\alpha$ profile, at a velocity range
between $-200$ and $-300$ km s$^{-1}$. This velocity coincides
with that of the high-velocity component in the forbidden line
profiles, peaking at $-300$ km s$^{-1}$ (Hamann 1994)
and considered to arise from a jet close to the star
(Eisl\"offel et al. 2000 for review). In addition,
the position angle of the displacement of 111$\pm$18$^\circ$
is coincident with that of a HH object
associated with this star, HH 32C (103$^\circ$ ---
Solf, B\"ohm \& Raga 1986).
We thus conclude that a jet close to the star
is responsible for the observed displacement.
One might think that a binary companion could also explain the observed displacement,
however, it is not likely since the emission component
in the H$\alpha$ profile does not show any displacement as observed
in known binaries (see \S 3.1).


Fig.~\ref{others} shows that the displacement has an angular scale of
70 milliarcsec, corresponding to $\sim$20 AU at the star.
The inclination of the extended jet is measured to be
$70 ^\circ$ from the plane of the sky
(Curiel et al. 1997); thus, the centroidal position of the
emission line region is presumably located
more than 60 AU apart from the star.
Position spectra in this object do not show a displacement corresponding to the counterflow,
as observed in RU Lup and CS Cha. Such results may be due to
obscuration of the flow by a circumstellar disc, as observed in forbidden
lines in many T Tauri stars.

\subsection{R CrA (Fig.~\ref{others})}
The positional displacement observed in H$\alpha$ emission differs from any other objects described above.
In 1996, the bluer component at $-$210 to 40 km s$^{-1}$ is displaced with an angular scale of 120 milliarcsec and at a position angle of $\sim$ 220$^{\circ}$, while the redder component at 110 to 600 km s$^{-1}$ does not show clear displacement. In 1999, the position of these components are displaced at position angles of $\sim$200$^{\circ}$ and $\sim$130$^{\circ}$ with angular scales of 100 and 70 milliarcsecs, respectively. Neither a stellar companion or a monopolar/bipolar outflow can explain these results.

Such results could be explained by a combination of two companions, or that of a stellar companion and an outflow. However, the former explanation is not likely for the following reason. Recent surveys show that young multiple systems are invariably hierarchical: i.e., with different orbital scales by a factor of at least 3 and typically more than 10 (e.g., Simon et al. 1995; Ghez et al. 1997a; K\"ohler et al 2000). This implies that two stellar companions cannot simply explain our results in R CrA, in which displacement in two directions show similar angular scales. A combination of a stellar companion and an outflow is thus more likely to explain the observed displacement.
Emission lines from outflows often exhibit only blueshifted components (e.g., Hirth, Mundt, \& Solf 1997): thus, a likely explanation is that the displacement in the bluer emission is due to an outflow, while that in the redder emission is due to a stellar companion.

Hillenbrand et al. (1992) estimate a primary stellar mass of 0.8 M$_\odot$ in this object, and our results provide a lower limit for the binary separation of 8 AU. These parameters provide the timescale of the orbital motion of more than 24 yr, much larger than the duration of our observations. This fact suggests that the observed time variation in the redshifted emission is due to the variation of the spectra in two binary components, as observed in T CrA (\S 3.2), rather than binary motion.

\begin{figure*}
  \begin{center}
    \leavevmode
\psfig{file=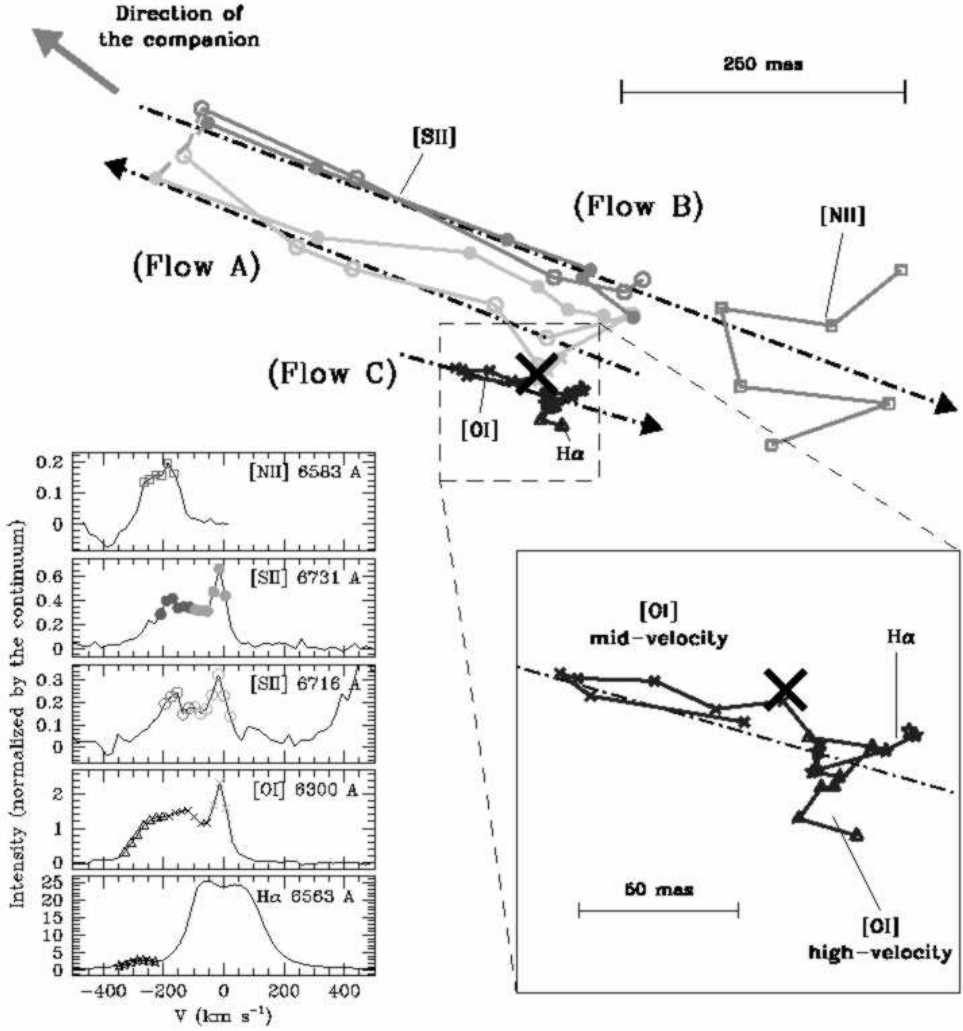,clip=,width=13 cm}
 \end{center}
\caption{Positional displacement in VV CrA. Contamination by the
continuum is removed as carried out for Fig.~\ref{CSCha}. Small open squares, filled and
open circles, and stars correspond to velocity components in
the profiles of \hbox{[N\,{\sc ii}]} 6583 {\AA},
\hbox{[S\,{\sc ii}]} 6731 and 6736 {\AA}, and H$\alpha$, respectively,
at the bottom-left of the figure. Small open triangles and crosses
are those for \hbox{[O\,{\sc i}]} 6300 {\AA} but different velocities
--- $-$330 to $-$200 km s$^{-1}$ and $-180$ to 10 km s$^{-1}$,
respectively. Different grey tones indicate the displacement
in flow A--C, respectively (see the text for detail).
A large cross shows the centroidal position of the continuum.
The dot-dashed arrows show the variation of the velocities
in the flows from low to high values.
The displacement in \hbox{[O\,{\sc i}]} 6363 {\AA} is not plotted, since it is consistent with that in \hbox{[O\,{\sc i}]} 6300 {\AA}, which has the same upper transition level. The intensity profiles in the bottom-left of the figure are normalized to the continuum.
}
  \label{VVCrA}
\end{figure*}

\subsection{VV CrA (Fig.~\ref{VVCrA})}

Positional displacement was observed in H$\alpha$ and several forbidden lines, and these are attributed to the following three flows: 
(1) flow A, which appears in the \hbox{[S\,{\sc ii}]} and \hbox{[O\,{\sc i}]} emission at velocities of $-$100 to 20 km s$^{-1}$; and
(2) flow B, which appears in the \hbox{[N\,{\sc ii}]} and \hbox{[S\,{\sc ii}]} emission at velocities of $-$270 to $-$100 km s$^{-1}$;
(3) flow C, which appears in the \hbox{[O\,{\sc i}]} and H$\alpha$ emission at velocities of $-$350 to $-$200 km s$^{-1}$. The position in the flow monotonically changes with velocity in the \hbox{[S\,{\sc ii}]} emission of flow A and B, and that in the \hbox{[O\,{\sc i}]} line in flow C.

The displacement in the lowest velocities of flow A is associated with the primary star shown in Fig.~\ref{VVCrA}, indicating that the flow is driven by the star at the centroidal continuum position. The \hbox{[S\,{\sc ii}]} profiles exhibit distinct components at high and low velocities, as observed in many other stars, and the latter component is located within 0.2 arcsec of the star, corresponding to spatial scales of $<$ 25 AU. This spatial scale is similar to those observed in other T Tauri stars (e.g., Hirth, Mundt, \& Solf 1997; Takami et al. 2001,2002), supporting the idea that the flow is driven by this PMS star. On the other hand, in flow B, the displacement at the lowest velocities lies in the north-east, in the direction of a companion with a separation of 2 arcsec (Graham 1992; Reipurth \& Zinnecker 1993; Ghez et al. 1997a). Thus, it is likely that flow B is driven by this companion, known as an ``infrared companion'' (Koresko, Herbst \& Leinert 1997). The [NII] emission in this flow is located further from this star than the [SII] lines, as observed in small-scale jets in many other T Tauri stars (Hirth, Mundt, \& Solf 1997), and such results support this conclusion.

Previous infrared observations by Graham (1992) and Ghez et al. (1997a) do not show
the presence of a young star associated with flow C. In this flow,
the \hbox{[O\,{\sc i}]} line has a width of a few hundred km s$^{-1}$,
suggesting the coexistence of shocks with different velocities. In addition,
the line with higher excitation energy, H$\alpha$, arises from the
highest velocities. These results can be attributed to a bow shock
in which the H$\alpha$ and \hbox{[O\,{\sc i}]} emission arise from its
apex and wakes, respectively. The extension of the \hbox{[O\,{\sc i}]} emission
suggest that the driving source lies in the west, in which a few $IRAS$ sources were identified by Wilking et al. (1992), although it is located far ($>$2 arcmin) from VV CrA. Thus, the driving source of flow C is not clear.

\section{H$\alpha$ profiles in binary components}
We can derive the intensity spectrum for each component of known binaries from the intensity and position spectra, the binary separation measured by infrared speckle techniques, and an assumed continuum flux ratio (see Bailey 1998a; Garcia, Thi\'ebaut, \& Bacon 1999). The centroidal position of the seeing in a PMS binary is described as follows:
\begin{eqnarray}
{\bf x_{\lambda}} &=& \frac{I_{\lambda \rm{(prim)}}}{I_{\lambda}}
                         {\bf x_{\rm{(prim)}}}
                   + \frac{I_{\lambda \rm{(sec)}}}{I_{\lambda}}
                         {\bf x_{\rm{(sec)}}}, \nonumber \\
                  & & \hspace{2cm} \rm{or} \nonumber \\
\Delta d_{\lambda}&=& \hspace{1.5cm}\frac{I_{\lambda \rm{(sec)}}}{I_{\lambda}} d,
\end{eqnarray}
where ${\bf x_{\lambda}}$ is the centroidal position of the seeing,
and ${\bf x_{\rm{(prim)}}}$ and
${\bf x_{\rm{(sec)}}}$ are positions of the primary
and the secondary stars, respectively; $I_{\lambda}$,
$I_{\lambda \rm{(prim)}}$, $I_{\lambda \rm{(sec)}}$ are intensities
of the whole system, the primary star, and the secondary star, respectively;
$\Delta d_{\lambda}$ is the separation between the primary star
and the centroidal position
of the seeing; $d$ is the binary
separation projected on the sky. Positional displacement in an emission/absorption feature
is measured relative to the centroidal position of the continuum emission,
thus its spatial scale is described as follows based on equation (2):
\begin{equation}
\Delta d_{\lambda}' = \Delta d_{\lambda} - \Delta d_{\rm{cont}}
                        = \left(  \frac{I_{\lambda \rm{(sec)}}}{I_{\lambda}} 
                        - \frac{I_{\rm{cont(sec)}}}{I_{\rm{cont}}} \right) d, \end{equation}
where $\Delta d_{\lambda}'$ is the observed displacement in a position spectrum;
$\Delta d_{\rm{cont}}$, $I_{\rm{cont}}$ and $I_{\rm{cont(sec)}}$ are
the distance to the centroidal continuum position from the primary star,
the total luminosity of the continuum, and that of the secondary
star, respectively. From this equation, the flux from each star
is obtained as follows:
\begin{eqnarray}
I_{\lambda \rm{(sec)}} &=& \left( \frac{I_{\rm{cont(sec)}}}{I_{\rm{cont}}} + \frac{\Delta d_{\lambda}'}{d} \right) I_{\lambda}
= \left( \frac{f}{1+f} + \frac{\Delta d_{\lambda}'}{d} \right) I_{\lambda},  \nonumber \\
I_{\lambda \rm{(prim)}} &=& I_{\lambda} - I_{\lambda \rm{(sec)}},
\end{eqnarray}
where $f$ is the continuum luminosity of the secondary
star relative to the primary.

\begin{figure*}
  \begin{center}
    \leavevmode
\psfig{file=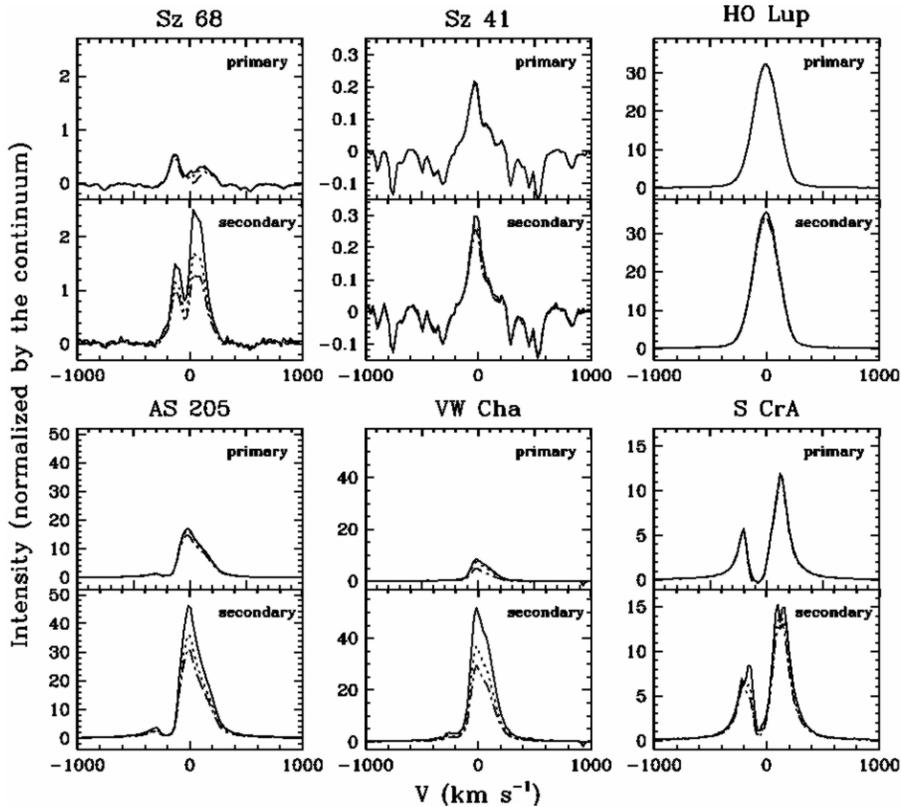,clip=,width=12 cm}
 \end{center}
\caption{Profiles of the H$\alpha$ emission for each binary component of known binaries Sz 68, Sz 41, HO Lup, AS 205, VW Cha, and S CrA. 
Solid, dot, and dot-dashed profiles were obtained by assuming the flux ratios of $f$=0.2, 0.4, and 0.8, respectively, at the continuum level. In this figure, the "primary" and "secondary" components are defined based on the continuum flux at 6500--6800 {\AA}, different from Fig.~\ref{knownbs2} and Table 2. The intensity scale is normalized to the continuum for each binary component.
}
  \label{profiles}
\end{figure*}

Fig.~\ref{profiles} shows the profiles of the H$\alpha$ emission obtained from equation (4) for six of our objects with known separation. Despite the uncertainty of the continuum flux ratio, the figure shows that the two binary components have strikingly similar profiles to each other --- similar linewidths and the absorption feature at similar velocities.
The figure also shows that the line-to-continuum ratio is higher at the secondary than the primary in all the objects. Although the profiles in HO Lup do not clearly show this tendency, the observed positional displacement lies in the direction of the fainter component (Reipurth \& Zinnecker 1993), agreeing with this conclusion.
These results also imply that the position of the H$\alpha$ feature in all the objects is primarily displaced in the direction of the component fainter at optical wavelengths. 

\section{Discussion}
\subsection{Detectability of close ($<$0.1 arcsec) binaries using spectro-astrometry}
Recent high-resolution techniques, such as infrared speckle observations and the $Hubble$ $Space$ $Telescope$, have revealed a difference in binary populations between star-forming regions. These observations show that the binary frequencies in low-mass star-forming regions are higher than that of solar-type field stars with the same range of binary separation (e.g., Ghez et al. 1993; Leinert et al. 1993; Ghez et al. 1997a). Alternatively, the frequency appears to be similar to that of the field stars in a massive star-forming region, the Orion Trapezium cluster (Prosser et al. 1994; Padgett et al. 1997; Petr et al. 1998).
Such a different binary population between regions could be caused by initial conditions, such as stellar density or cloud temperature (see Mathieu et al. 2000 for review).

However, the traditional observing techniques have not been able to cover the complete range of binary separations in various star-forming regions. Direct imaging and infrared speckle have allowed detection of PMS binaries down to separations of $\sim$100 milliarcsecond, corresponding to $\sim$15 AU with periods of 50 years or more in the nearest star-forming regions. On the other hand, spectroscopic measurement has been capable of detecting binaries with periods of a few days and separations of well under an AU. These techniques cannot cover the intermediate separations (0.1--10 AU), leading to significant incompleteness. Lunar occultations have explored this gap in some cases, however, this technique can measure the separation only in one-dimension in the sky, and applies for a limited number of star-forming regions as well as relying on chance occultations.

Spectro-astrometric observations by Bailey (1998b) and this work suggest that this technique is capable of discovering PMS binaries with these intermediate separations. These observations detect 10 out of 12 known binaries with separations of 0.07--1.5 arcsec, and a few new binary candidates which have not been detected using other techniques. The observed angular scales in known binaries (0.07--0.7 times as large as the binary separation --- \S 4) and positional accuracy achieved (down to 1 milliarcsec) suggest that the technique should be able to detect binaries with separations down to $\sim$10 milliarcsec at the 3-$\sigma$ level. This detection limit is comparable to that of lunar occultations (see Richichi et al. 1994; Simon et al. 1995), although spectro-astrometry can easily achieve better limits by detecting a larger number of photons (see equation 1), and applies for objects with any coordinate and binary position angles. 

\begin{figure}
  \begin{center}
    \leavevmode
\psfig{file=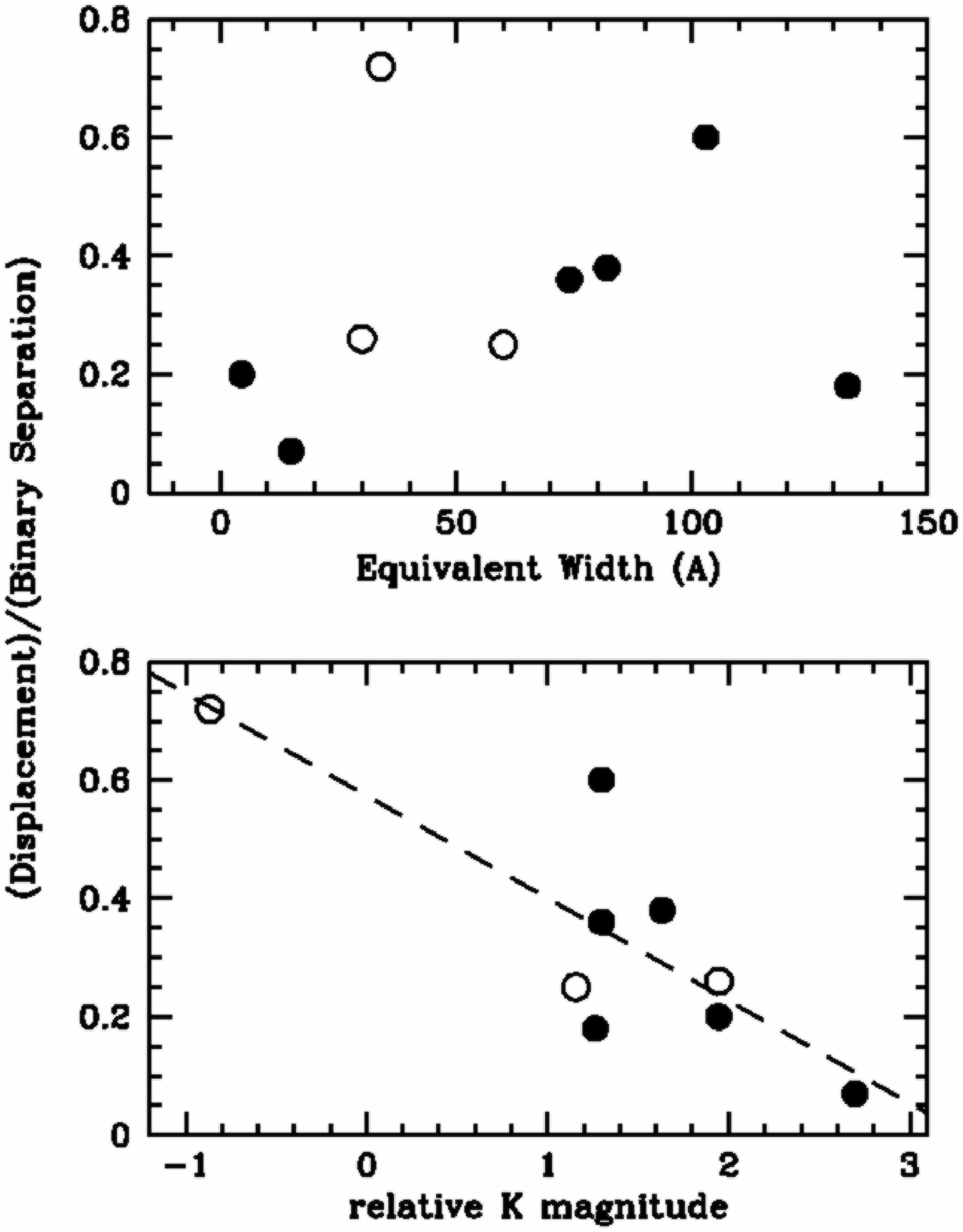,clip=,width=7.5 cm}
 \end{center}
\caption{Relation between the angular scale of the displacement and H$\alpha$ equivalent width of the whole system (upper), and relative $K$ magnitude between binary components (lower). The open and filled circles show the results for Herbig Ae/Be stars and classical T Tauri stars, respectively. The regression line (dashed) is shown in the lower figure. Data for the displacement and H$\alpha$ equivalent widths are obtained from Bailey (1998b) and this work, while the relative $K$ magnitudes are obtained from previous infrared speckle observations by Ghez et al. (1993), Ghez et al. (1997a), and Leinert et al. (1997).
}
  \label{correlation}
\end{figure}

Equation (3) shows that the detectability of binaries using this technique depends on the line and continuum flux ratios between binary components. In addition, the position spectra in H$\alpha$ emission show different shapes from the intensity profiles (see Fig.~1), indicating that the detectability also depends on a slight difference in the line profiles for the primary and secondary stars. These facts may be susceptible to more complicated selection effects than those of direct imaging, infrared speckle, and lunar occultations. To investigate this effect, we plot the relation between the angular scale of the displacement, H$\alpha$ equivalent width of the whole system, and relative $K$ magnitude between binary components in Fig.~\ref{correlation}. The figure shows that the displacement has no correlation with the H$\alpha$ equivalent width, while there is a clear correlation with the relative $K$ magnitude. These results suggest that our technique has a selection effect similar to those of infrared speckle (Ghez et al. 1993) and lunar occultations (Simon et al. 1995); i.e., dependent on the relative $K$ magnitude and independent of the line flux.
The latter correlation between the displacement and relative $K$ magnitude can be explained as follows: if the secondary star has a bright stellar/disc luminosity, it is likely to have bright H$\alpha$ emission (see Cabrit et al 1990) and bright $K$ band emission, thereby providing a small difference in the $K$ magnitude from that of the primary and a large displacement in H$\alpha$ emission towards the secondary.

\begin{figure}
  \begin{center}
    \leavevmode
\psfig{file=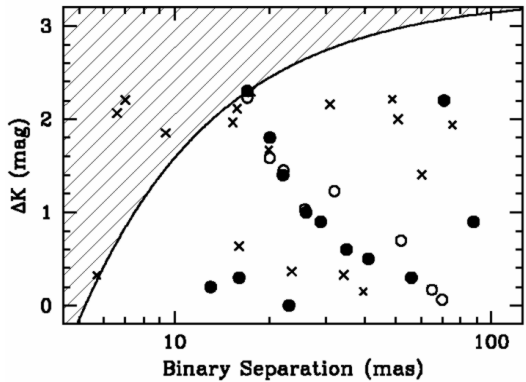,clip=,width=7 cm}
 \end{center}
\caption{The detection limit of PMS binaries using spectro-astrometry. The hatched area shows the parameter space in which we can{\it not} detect binaries exceeding the 3-$\sigma$ level. This limit is estimated assuming a positional uncertainty of 1 mas in the position spectra. The filled and open circles, and crosses show the binaries discovered by lunar occultations by Simon et al. (1995), Richichi et al. (1994), and Richichi et al. (1996, 1997), respectively.
}
  \label{detectionlimit}
\end{figure}

From the regression line in Fig.~\ref{correlation}, we estimate the detection limit of binary separation using our technique as a function of the relative $K$ magnitude (Fig.~\ref{detectionlimit}). The same figure also shows binary parameters discovered by lunar occultations with separations of less than 0".1 (Richichi et al. 1994,1996,1997; Simon et al. 1995). The figure indicates that we could detect 90 \% of binaries discovered by lunar occultations, and our detection limit appears to be slightly better than that of the lunar occultation method for both binary separation and relative $K$ magnitude. Such a high sensitivity for relative $K$ magnitude is also followed by a discovery of a companion candidate in T CrA, which Ghez et al. (1997a) failed to detect despite a sufficient separation ($>$0.14 arcsec --- see \S 3.2).

In two separated studies, we failed to detect the companions in T Tau (Bailey 1998b) and SR 20 (this work), which were detected by Ghez et al. (1993) using infrared speckle. The non-detection for the former object could be due to the fact that the secondary star is an  ``infrared companion'' and faint at optical wavelengths (Koresko et al. 1997). Such companions form only $\sim$10\% (White \& Ghez 2001) or less (Ghez et al. 1997a; K\"ohler et al. 2000) of the entire PMS population. Another possible explanation for non-detection is that H$\alpha$ emission is faint or absent at the companion of T Tau and SR 20, thereby not allowing the positional displacement towards the secondary. Again, such binary systems may not significantly affect the entire population for PMS binaries: recent observations have shown that companions of classical T Tauri stars (CTTSs) are usually CTTSs, and a combination of a CTTS and a weak-line T Tauri star (WTTS) is rare ($\la$20\% --- Prato \& Simon 1997; Duch\^ene et al. 1999).

All the results described above suggest that we can improve the previous statistics of PMS binary populations using spectro-astrometry. Note that most of the known binaries in our sample have bright H$\alpha$ emission ($EW_{\rm{H\alpha}}$$>$10 \AA), implying that the detection of PMS binaries using our technique could be efficient only for such objects as CTTSs. Indeed, equation (3) suggests that the displacement could be shorter for binaries with small line-to-continuum ratios. Although Fig.~\ref{correlation} does not show such a tendency, it could appear if we included the objects with fainter H$\alpha$ emission. This fact implies the detectability of binaries using our technique could be worse for objects such as WTTSs.

\subsection{Binary frequency in our sample of the nearest southern star-forming regions}
The observed targets shown in Table 1 include 22 PMS stars in nearby (130--200 pc) star-forming regions: including 7 known binaries with separations of 0.07--1.5 arcsec (\S 3.1), and two ``spectro-astrometric'' binaries in T CrA (\S 3.2) and R CrA (\S 3.6). Among these binaries, R CrA may have a separation of less than 0.1 arcsec, corresponding to $\la$15 AU. In contrast, lunar occultations surveys by Simon et al. (1995) detected 8 binaries out of 47 PMS stars in Taurus and 5 out of 35 PMS stars in Ophiuchus at these short ($\la$15 AU) separations. Is the binary frequency at these separations smaller in our sample than the Taurus and Ophiuchus star-forming regions?

To investigate this, we determine the binary frequency of our sample. We exclude Sz Cha from this statistical study due to a remarkably low detection limit (13 mas). The objects in which a companion has not been detected have a typical uncertainty in their measured displacement of 3 milliarcsec. Given that the typical observed angular scale per binary separation is 0.3 for H$\alpha$ emission (see \S 3.1 and Table 1), this implies that a companion has not been detected down to 30 milliarcsec (3$\sigma$), equivalent with $\sim$5 AU in the observed star-forming regions. Table 3 shows binary populations in our sample with two different separation ranges (15--300 AU and 5--15 AU), together with those in various star-forming regions and field stars. An accurate binary separation is not known for R CrA, and this fact is included in the uncertainty of the tabulated values.
The table shows that our sample has a larger binary frequency for separations of 15--300 AU than that in field stars. Such a result is consistent with those derived in other low-mass star-forming regions. On the other hand, all the frequencies tabulated for the shorter separations have large statistical uncertainties, implying that the provided separation range is insufficient to investigate the difference of binary frequencies between star-forming regions. Observations of 60 stars with an uncertainty of 1 mas in their position spectra could allow us to investigate it in detail: we could detect 8$\pm$3 companions with these separation ranges if the binaries in the observed regions had the same number distribution of separations as that in solar-type main-sequence stars observed by Duquennoy \& Mayor (1991).

\begin{table}
 \centering
 \begin{minipage}{80mm}
 \caption{Binary population in star-forming regions}
 \label{tab:table}
 \begin{tabular}{@{}lcl@{}}
\multicolumn{3}{c}{--- separation = 15--300 AU\footnote{assuming the distance of all the star-forming regions of 150 pc for simplicity} ---}\\[5pt]
region & binary fraction & reference \\[10pt]
Taurus      & 0.37$\pm$0.09 & Ghez et al. ~~(1993)\\
            & 0.32$\pm$0.08 & Simon et al. ~(1995)\\
Ophiuchus   & 0.14$\pm$0.06 & Simon et al. ~(1995)\\
Oph-Sco     & 0.29$\pm$0.12 & Ghez et al. ~~(1993)\\
Sco-Cen     & 0.26$\pm$0.05 & K\"ohler et al. (2000)\\
Cha/Lup/CrA & 0.31$\pm$0.08 & Ghez et al. ~~(1997a)\\
field stars & ~~0.16$\pm$0.04\footnote{adopting the factor by Leinert et al. (1993) to convert the orbital period to the separation} & DM91\footnote{Duquennoy \& Mayor (1991)} \\ [2pt]
our sample in\\
~southern SFRs & 0.41$\pm$0.16 & this work\\[10pt]
\multicolumn{3}{c}{--- separation = 5--15 AU$^a$ ---}\\[5pt]
region & binary fraction & reference \\[10pt]
Taurus      & 0.06$\pm$0.04 & Simon et al. (1995)\\
Ophiuchus   & 0.06$\pm$0.04 & Simon et al. (1995)\\
field stars & ~~0.06$\pm$0.02$^b$ & DM91$^c$ \\ [2pt]
our sample in\\
~southern SFRs& 0.05$\pm$0.05 & this work\\
\end{tabular}
\end{minipage}
\end{table}

\subsection{Implications for the binary formation mechanism}
Approximately two-thirds of all solar-type main-sequence stars are members of multiple systems (Duquennoy \& Mayor 1991), and it may be even higher in low-mass star-forming regions (see Mathieu et al. 2000).
Despite the apparent ubiquity of binaries, astronomers have not yet identified the primary mechanism by which they form. Possible formation mechanisms include (see Hartigan, Strom, \& Strom 1994a; Ghez, White, \& Simon 1997b): (1) fragmentation of a collapsing molecular core, (2) gravitational instability in an equilibrium disc, and (3) dynamical capture. The secondary mass distribution in field stars is remarkably similar to the mass function of primary stars with low-to-high masses (Fischer \& Marcy 1992; Duquennoy \& Mayor 1991; Abt, Gomez, \& Levy 1990), supporting protostellar ``capture'' as the mechanism by which binaries form. On the other hand, stellar ages, infrared excess, and emission line features in PMS binaries indicate that these are at the same evolutionary stage, supporting the scenario of core fragmentation (e.g., Hartigan et al. 1994; Ghez et al. 1997b; Prato \& Simon 1997; Duch\^{e}ne et al. 1999; White \& Ghez 2001). 

Our results in \S 4 show that the two binary components have strikingly similar H$\alpha$ profiles to each other. Such results contrast to a rich variety of profiles among PMS stars (see Reipurth, Pedrosa, \& Lago 1996). The variety of these profiles is due to interplay between the accretion flow and a wind closest to the source: thus, our results suggest that the two binary components tend to have similar flow geometry and velocity for mass accretion/ejection. Such results could exclude the scenario of dynamical capture and disc instability for binary formation, since (1) capture would provide a rich variety of profiles even between binary components, and (2) disc instability could provide different evolutionary stages between binary components, thereby providing different inflowing/outflowing activities. Our results thus support the scenario of core fragmentation for binary formation, as well as other studies for PMS binaries

Accretion flows and/or winds in YSOs are likely to have non-spherical geometry (Shu et al. 2000; K\"onigl et al. 2000; Najita et al. 2000): thus, the similarity of the H$\alpha$ profiles observed in binary components would also require the condition that the axes of the inflows/outflows are aligned with each other. Such a condition is consistent with a previous polarimetric study by Monin, M\'enard, \& Duch\^ene (1998), who show that the position angles of polarization are preferentially parallel in PMS binaries. On the other hand, Hale (1994) shows that the orbital and equatorial planes appear to be randomly distributed in main-sequence binaries with separations more than 30--40 AU. Such a difference between PMS and main-sequence binaries could result from different formation mechanisms, or different initial conditions of the protostellar core. In the latter case, the results in PMS binaries could come from the pre-collapse material all rotating around a single axis, while those in main-sequence binaries could originate from an the initial cloud core in which the angular momentum distribution has strong spatial variations (see the discussion of Bate et al. 2000).

\subsection{Physical Conditions of Jets down to AU Scales}
The results for the H$\alpha$ outflows obtained by Bailey (1998b), Takami et al. (2001) and this work are summarized in Table 4. Among these objects, three of them are known as active PMS stars: Z CMa and AS 353A are associated with extended jets (e.g., Poetzel, Mundt, \& Ray 1989; Solf, B\"ohm, \& Raga 1986) and RU Lup shows extremely high accretion luminosity (3--6 L$_\odot$ --- Lamzin et al. 1996).
Such high activities have not been explicitly reported for CS Cha, although the object shows high variability in H$\alpha$ emission:- measured equivalent widths range from 13 \AA  ~(Herbig \& Bell 1988) to 54 \AA  ~(Reipurth, Pedrosa, \& Lago 1996) and 60 \AA ~(this work), and the observed line profile in Fig.~\ref{CSCha} show a different shape from that observed by Reipurth et al. (1996).

\begin{table}
 \centering
 \begin{minipage}{80mm}
 \caption{Summary of H$\alpha$ outflows}
\begin{tabular}{@{}lccc@{}}
object  & projected scale & velocity$^a$ & $EW_{H\alpha}$\footnote{sum for the blueshifted and redshifted emission in which positional displacement is observed}\\
& (AU)& (km s$^{-1}$)  & (\AA)\\[10pt]
RU Lup &$\sim$5\footnote{larger than that in Takami et al. (2001) due to the updated distance (Wichmann et al. 1998)}
&$-200$$\leq$$v$$\leq$~$-$50&29.7\\ 
&&~~$~200$$\leq$$v$$\leq$~300\\ 
CS Cha &$\sim$1.5&$-100$$\leq$$v$$\leq$~$-$20&28.3\\
&&~~$~100$$\leq$$v$$\leq$~250\\ 
Z CMa   & $>70$\footnote{adopting the distance of 1150 pc provided by Clari\'a (1974)}& $-700$$\leq$$v$$\leq$~$-$50&---\footnote{equivalent widths cannot be derived since the displacement is overlapped with the absorption feature}\\
AS 353A &      $>20$    & $-300$$\leq$$v$$\leq$$-$200&---$^e$\\

\end{tabular}
\end{minipage}
\end{table}

The velocities of blueshifted H$\alpha$ outflows in RU Lup, Z CMa, AS 353A coincide with those of the high-velocity components in forbidden lines: $-$100 to $-$250 km s$^{-1}$ in RU Lup (Takami et al. 2001); $-600$ to $-800$ km s$^{-1}$ in Z CMa (Corcoran \& Ray 1997); $-200$ to $-350$ km s$^{-1}$ in AS 353A (Hamann 1994).
This forbidden line emission is considered to arise from jets close to the star (see Eis\"offel et al. 2000, and references therein): thus, this coincidence of velocities suggests that the H$\alpha$ emission in these objects also arises from jets close to the source.
Although the H$\alpha$ emission in PMS stars are often considered to arise from the magnetospheric accretion columns, these results are not surprising for the following reasons.
First, the idealized magnetospheric accretion models by Muzerolle et al. (1998,2001) failed to reproduce permitted line profiles in active PMS stars, suggesting that the presence of another contributor to the emission. In addition, dozens of YSOs exhibit thermal radio emission, which is likely due to the ionized gas in jets close to the source (see Anglada 1996). From the H$\alpha$ luminosities in RU Lup and CS Cha, we can derive upper limits of the thermal radio luminosities of 0.1--0.4 mJy kpc$^2$ at 3.6 cm (see Appendix A), consistent with those observed in other Class II objects (see Beltr\'an et al. 2001)\footnote{We assume that the magnetospheric accretion columns do not significantly contribute to the radio flux due to its large optical thickness and small size, or obscuration by an optically thick jet/wind. The latter explanation is often used for the fact that WTTSs often show nonthermal radio emission from the magnetosphere, while it is usually absent in CTTSs and Class I YSOs (see, e.g., Andr\'e et al. 1992; Martin 1996).}.

H$\alpha$ emission-line regions in Z CMa and AS 353A show spatial scales more than 50 AU (see Table 4 and \S 3.5), and there is growing evidence that the jets at these scales are heated by shocks in the internal working surfaces.
Recent high-resolution imaging of PMS stars exhibit knotty, bubbly, and bow-shape structures in jets at these scales (e.g., Kepner et al. 1993; Bacciotti et al. 2000; Dougados et al. 2000).
Lavalley-Forquet et al. (2000) carried out a detailed analysis of line ratios in DG Tau, and reach the conclusion that the jet is heated by shock, and not ambipolar diffusion nor turbulent viscosity in the mixing layer.
Their conclusion is supported by Takami et al. (2002), who discover \hbox{He\,{\sc i}} 1.083 $\mu$m emission in the same jet: only shocks could explain the presence of such a high-excitation ($\sim$20 eV) line.

On the other hand, the same heating mechanism cannot simply explain the following results in RU Lup and CS Cha. H$\alpha$ emission in these objects show spatial scales of 1--5 AU, much smaller than the distance between internal working surfaces observed in PMS jets ($\ga$100 AU --- see, e.g., Ray et al. 1996; Dougados et al. 2000). In addition, their line profiles do not exhibit a peak at the velocity of the jets, and indicate that H$\alpha$ emission in these jets is continuously distributed from the accelerating region. These results suggest the presence of another heating mechanism in the inner ($<$10 AU) region, which could be (1) X-ray heating (Shang et al. 2002); (2) MHD-wave (e.g., Hartmann, Edwards, \& Avrett 1982); (3) ambipolar diffusion (Safier 1993; Garcia et al. 2001); (4) turbulent dissipation in a viscous mixing layer (Binette et al. 1999); or (5) UV photoionization. Among these mechanisms, ambipolar diffusion and viscous mixing layer models provide a lower ionization fraction than shocks, as indicated by the predicted [NII]/[OI] line ratios (see Lavalley-Fouquet et al. 2000). Such a prediction is inconsistent with the spatial distributions of H$\alpha$ and [OI] emission in RU Lup, which indicates a higher ionization fraction in the inner region than for the outer region heated by shocks. PMS stars often exhibit UV continuum excess as a result of mass accretion (see e.g., Calvet, Hartmann, \& Strom 2000), although this could also be rejected for the following reason. The hot continuum emission of RU Lup at $>$1200 \AA ~is approximately fitted by a blackbody with a temperature of 6.8--7.8$\times$10$^3$ K and a luminosity of 2.7--6 L$_\odot$ (Lamzin et al. 1996), and a flux of ionizing photons of $10^{37}$--$10^{38}$ s$^{-1}$ from this region could sustain only 10$^{-4}$--10$^{-3}$\% of the recombination rate corresponding to the observed H$\alpha$ luminosity. Thus, one of the remaining two mechanisms, X-ray or MHD-wave heating, would be more likely for the heating mechanism in this region. The same mechanisms are also proposed for magnetospheric accretion columns, which are responsible for Balmer and other permitted lines in many PMS stars (Lamzin et al. 1996; Calvet \& Gullbring 1998).

From the spatial scales, flow velocities, and H$\alpha$ luminosities, we can estimate the typical hydrogen number densities and opening angles in these AU-scale outflows.  The details of the calculation is described in Appendix B. Throughout we assume the following: (1) an electron temperature of 10$^4$ K, (2) viewing angle of 45$^{\circ}$, (3) $\dot{M}_{out} = 0.1 \times \dot{M}_{acc}$, where $\dot{M}_{out}$ and $\dot{M}_{acc}$ are the mass loss rate and mass accretion rate, respectively, and (4) the flow has axisymmetric geometry and zero width at its origin. The first assumption is based on models which include X-ray heating (Shang et al. 2002) or the temperature of magnetospheric accretion columns (e.g., Muzerolle et al. 2001), in which the presence of MHD-wave heating is suggested by Calvet \& Gullbring (1998). The second assumption is likely to be a good approximation for order-estimates, since viewing angles close to pole-on and edge-on would not provide the observed displacements and velocities, respectively. The third assumption is based on Calvet (1997), who revised the estimate from forbidden line and accretion luminosities by Hartigan et al. (1995).
For mass accretion rates in RU Lup and CS Cha, we adopt 7.7$\times$10$^7$ and 1.6$\times$10$^7$ M$_\odot$ yr$^{-1}$, respectively, obtained by John-Krull, Valenti, \& Linsky (2000).

\begin{figure}
  \begin{center}
    \leavevmode
\psfig{file=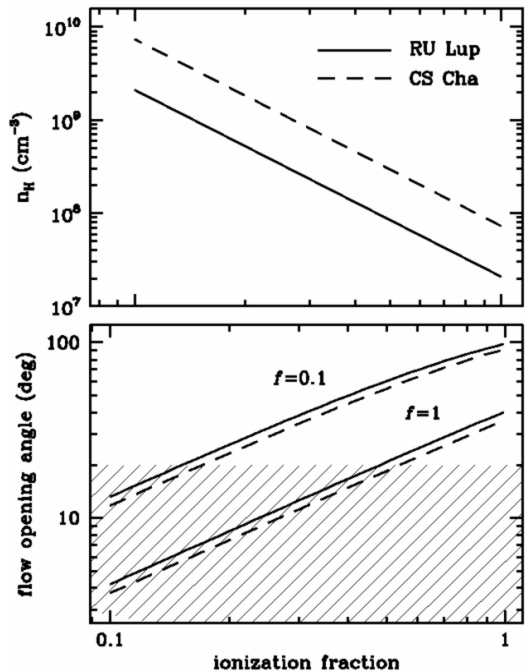,clip=,width=7 cm}
 \end{center}
\caption{Mean hydrogen density and flow opening angle as a function of the ionization fraction. The estimated opening factor depends on the volume filling factor in the flow, which are assumed to be 0.1 and 1. The hached area indicates unlikely opening angles (see the text).}
  \label{outflow}
\end{figure}

Fig.~\ref{outflow} shows the derived hydrogen number densities and opening angles in RU Lup and CS Cha, as a function of the assumed ionization fraction. The upper figure shows that the hydrogen number density is $\ga 10^7$ cm$^{-3}$ for any case, and larger for lower ionization fractions. Previous imaging observations have shown that opening angles of jets are likely to be $\ga$20$^\circ$ within $\sim$100 AU of the stars (see Eis\"offel et al. 2000, and references therein; see also Bacciotti et al. 2000), and the lower figure shows that this fact gives lower limits to the ionization fraction of 0.2--0.5 for the assumed volume filling factors of 0.1--1, providing the upper limits of n$_H$$\sim$10$^9$ cm$^{-3}$.

\subsection{Disc gap/hole in CS Cha and RU Lup --- implications for planet formation}
Recent radial velocity surveys have revealed a number of extrasolar giant planets (see, e.g., Marcy, Cochran, \& Mayor 2000). 
These observations show a population minimum between stellar-substellar  companions and extrasolar giant planets (``the brown dwarf desert'' --- Marcy, Cochran, \& Mayor 2000; see also Halbwachs et al. 2000), suggesting the presence of another formation mechanism for the planetary-mass objects. The proposed mechanisms include: (1) core accretion via accumulation of dust grains, which is generally accepted and requires 10$^6$--10$^7$ yrs for planet formation (e.g., Pollack et al. 1996), and (2) disc instability, which could form a gas giant planet only in a few hundred years (Boss 2000).

Model calculations have shown that a young planet could tidally interact with a circumstellar disc, opening up a gap or hole at its orbital scale (see Lubow \& Artymowicz 2000; Takeuchi, Miyama, \& Lin 1996). Even the most recent high-resolution techniques cannot resolve such AU-scale structures in the nearest star-forming regions, while infrared spectral energy distributions (SEDs) in several young stars exhibit dips at the mid-infrared, suggesting the presence of such gaps/holes (e.g., Strom et al. 1989; Marsh \& Mahoney 1992,1993). Recent studies by Boss \& Yorke (1993,1996) show that the SEDs cannot provide convincing proof for the presence of a disc gap or hole as these dips could also be caused by a combination of vertical temperature structure and changes in dust opacity.

On the other hand, our results for RU Lup (Takami et al. 2001) and CS Cha (this work) are providing evidence for the presence of disc gaps/holes at AU scales. These objects exhibit the presence of H$\alpha$ bipolar jets at AU scales, and the detection of the redshifted flows is explained by a disc gap/hole allowing this flow to be seen (Takami et al. 2001; \S 3.4).
Our position spectra would also be capable of discovering a stellar companion. No evidence for companion is seen, suggesting that the gap and hole in these objects could be induced by unseen companions.

The SEDs suggest that the outer radii of the gap and hole correspond to a temperature of $\sim$200 K, coincident with the ice condensation temperature (160--200 K --- see Marsh \& Mahoney 1992; Boss 1995).
Marsh \& Mahoney (1992,1993) also obtained similar temperatures from mid-infrared dips in GK Tau and HK Tau SEDs for which stellar companions have not been detected.
In contrast, the gaps/holes induced by a stellar companion show a wide range of temperatures: Guilloteau, Dutrey, \& Simon (1999) provide an inner temperature of the GG Tau ring of $\sim$ 30 K, while results by Jensen \& Mathieu (1997) suggest temperatures up to $\sim$600 K in the gaps/holes towards spectroscopic binaries. Marsh \& Mahoney (1992,1993) suggest that such results could be explained if a gas-giant planet formed at the ice condensation radius and clearing up the innerside of the circumstellar disc by tidal interaction.
Indeed, theoretical work shows that the formation time scales of the gas-giant planets are likely to be shorter at the ice condensation radius or outer region than the inner region: the presence of water ice could enhance the surface density of the solid material by a factor of 3--4 (e.g., Hayashi 1981), providing a shorter timescale for the planetesimal accretion (Lissauer 1987). Furthermore, water vapor produced in the inner region could condensate at the ice condensation radius, enhancing the surface density of ice by a factor of $\sim$70 (Stevenson \& Lunine 1988), thereby allowing the ice condensation radius to be a preferential location for formation of giant planets.

RU Lup, CS Cha, GK Tau, and HK Tau have stellar ages of 1--3$\times$10$^6$ yr (see Palla \& Stahler 1999 for the evolutionary tracks, and Lamzin et al. 1996, Gauvin \& Strom 1992, and  Marsh \& Mahoney 1993 for their effective temperatures and luminosities), implying that the young planets could be formed within this timescale.
This result contrasts with the detailed simulations by Pollack et al. (1996), who suggest that $\sim$5$\times$10$^6$ yr is required for the formation of a Jupiter in the solar nebula.
This age problem could be overcome if the initial surface density is higher than their standard model (10 g cm$^{-2}$ for solid material at the scale of the Jovian orbit). These authors show that the timescale for the formation of giant planets appears to be extremely sensitive to the initial surface density of the protoplanetary disc: e.g., increasing the density by a factor of 2 could reduce the timescale by a factor of 30 for the formation of Jupiter. Such higher initial surface densities are likely to provide higher planetary mass than Jupiter (Lissauer 1995; Pollack et al. 1996), consistent with the discovery of a number of extrasolar planets with such masses (see Marcy \& Butler 2000; Jorissen, Mayor, \& Udry 2001). We thus conclude that the disc gaps and holes observed in RU Lup and CS Cha are consistent with the scenario that these are induced by a gas-giant planet formed at the ice condensation radius.



\section{Conclusions}
We present spectro-astrometric observations for 28 southern PMS stars. The structures detected in the position spectra include: (1) almost all the known binary companions in our sample (Sz 68, Sz 41, HO Lup, VW Cha, S CrA, and AS 205), (2) companion candidates which have not been detected by infrared speckle techniques (T CrA, MWC 300), (3) monopolar and bipolar jets (AS 353A, CS Cha), 
(4) a combination of a jet and stellar companion (R CrA), and (5) a combination of jets and a bow shock (VV CrA).

We succeeded to detect almost all the known binaries in our sample (6 out of 7) with separations of 0.1--1.5 arcsecs. The position of H$\alpha$ emission in these objects are displaced primarily in the direction of the faint component, and their angular scales are consistent with those obtained by Bailey (1998b): 0.07--0.7 times as large as the binary separation. 
The displacement in the other features is much shorter than those in the H$\alpha$ emission (0.02--0.05 times as large as binary separation) due to larger contribution of the continuum flux. All the observed displacements lie in the direction of the binary system, and these are explained by different spectra between the two binary components. Among the ``spectro-astrometric'' binary candidates, we confirm the time variation of the position angle in T CrA, consistent with binary motion.

From results in known binaries to date, we conclude that this technique should be able to detect binaries with separations down to $\sim$10 milliarcsec. This detection limit is comparable to that of lunar occultations, although our technique can easily achieve better limits by observing a larger number of photons (\S 2), and is applicable for objects with any coordinate and binary position angles.
The H$\alpha$ profiles obtained for the components of the known binaries are strikingly similar to each other --- similar linewidths and absorption features at similar velocities, results which support the scenario of core fragmentation for the mechanism of binary formation. 

The H$\alpha$ outflows observed by Bailey (1998b), Takami et al. (2001), and this work exhibit velocities coincident with the high-velocity components of the forbidden lines, suggesting that these are associated with jets close to the source. Among these objects, Z CMa and AS 353A show spatial scales more than 50 AU, and these are likely to be excited by shocks in the internal working surfaces. On the other hand, bipolar outflows in RU Lup and CS Cha exhibit much shorter spatial scales (1--5 AU), and the spatial scale and line profile in these objects suggest the presence of a heating mechanism other than shocks: e.g., X-ray or MHD-wave heating. From the observed spatial scales, flow velocities, and H$\alpha$ luminosities of these flows, we estimate typical hydrogen number densities $\ga 10^7$ cm$^{-3}$.

Our results in RU Lup (Takami et al. 2001) and CS Cha (this work) suggest the presence of disc gaps/holes at AU scales, consistent with the presence of a mid-infrared dip in their SEDs. The SEDs suggest that the outer radii of these gap and hole have temperatures of $\sim$200 K, coincident with the ice condensation temperature (160--200 K). The constant temperature in these objects and a few more disc gap/hole candidates contrasts with the wide range for those in PMS binaries (30--600 K). The presence of such disc gaps/holes could be explained if a gas-giant planet formed at the ice condensation radius, clearing up the innerside of the circumstellar disc by tidal interaction. \vspace{0.5cm}\\

\begin{acknowledgements}
We thank the staff at the Anglo-Australian Observatory
for their support during the observations. We are grateful to
the referee (Dr. F. Comeron) for his careful review and providing useful comments.
We acknowledge
the data analysis facilities provided by the Starlink Project which
is run by CCLRC on behalf of PPARC.
MT thanks PPARC for support through a PDRA.
This research has made use of Simbad database, operated at CDS. Strasbourg, France, and of the NASA's Astrophysics Data System Abstract Service.
\end{acknowledgements}

\appendix
\section[]{Expected luminosities for thermal radio emission}
The fluxes of H$\alpha$ and free-free emission per unit volume are described as follows (see Osterbrock 1989; Reynolds 1986):
\begin{eqnarray}
4 \pi j_{H\alpha} &=& 3.5 \times 10^{-25} n_p n_e ~~~~~~~~\rm{erg~cm^{-3}~s^{-1}}\\
4 \pi j_{\nu} &=& 3.3 \times 10^{-38} n_p n_e \nu^{-0.1} ~ \rm{erg~cm^{-3}~s^{-1}~Hz^{-1}}
\end{eqnarray}
where $n_p$ and $n_e$ are the proton and electron densities, respectively; $\nu$ is the observed frequency for free-free emission. Throughout we assume an electron temperature of 10$^4$ K, plausible for either shock heating (see, e.g., Hartigan, Morse, \&  Raymond 1994b), ambipolar diffusion (Safier 1993; Garcia et al. 2001), or UV photoionization (e.g., Osterbrock 1989) as a heating mechanism. We also assume that the H$\alpha$ emission is produced via recombination, and the contribution from collisional excitation is negligible. Indeed, the transition level of H$\alpha$ is too high to be excited thermally at the assumed temperature (Osterbrock 1989). Collisional excitation of H$\alpha$ could occur in fast shocks, although the total H$\alpha$ flux in these shocks may still be dominated by the ``recombination plateau'', which has a temperature of $\sim$10$^4$ K and a much larger volume (e.g., by a factor of 10$^3$) than the region in which the line is collisionally excited (see Figs 2--3 of Bacciotti \& Eisl\"offel 1999).

Equations (A1) and (A2) show that free-free emission is proportional to that of H$\alpha$ emission,
\begin{eqnarray}
j_{3.6 cm} &=& 9.4 \times 10^{-15} j_{H\alpha} ~~~~~~~~~~\rm{~Hz^{-1}}
\end{eqnarray}
Table A.1 summarizes the observed parameters, and the estimated luminosities for free-free emission for the optically thin case.
Note that the derived free-free luminosities are upper limits, since the radio thermal emission in YSOs are often optically thick (see Beltr\'an et al. 2001, and references therein).

\begin{table}
 \begin{minipage}{80mm}
 \caption{Estimated radio luminosities}
\begin{tabular}{@{}lccccc@{}}  
object & $R$\footnote{Herbig \& Bell (1988)} & $EW$\footnote{sum for the blueshifted and redshifted emission in which positional displacement is observed} & $A_V$ & $L_{H\alpha}$\footnote{adopting the distances obtained by Wichmann et al. (1998)} & $S_{3.6cm}d^2$ \\ 
& (mag) & ({\AA}) & (mag) & (L$_\odot$)&(mJy kpc$^{2}$)\\ [10pt]
RU Lup & ~9.99 & 29.7 & 1.28\footnote{Hughes et al. (1994)} & 1.4$\times$10$^{-2}$& 0.42\\
CS Cha & 10.84 & 28.3 & 0.85\footnote{Gauvin \& Strom (1992)}   & 3.6$\times$10$^{-3}$& 0.11\\  
\end{tabular}\\
\end{minipage}
\end{table}

\section[]{Hydrogen density and opening angle of the H$\alpha$ bipolar outflow}

The total mass of the H$\alpha$ emission line region and the H$\alpha$ luminosity are described as follows:
\begin{eqnarray}
M_{out} &\sim& \int m_p n_H ~dV = \dot{M}_{out} l_{out} / v_{out}\\
L_{H\alpha} &=& 3.5 \times 10^{-25} \int n_p n_e~dV~~~\rm{erg~s^{-1}}
\end{eqnarray}
where $m_p$ and $n_H$ are the proton mass and the hydrogen number density, respectively; $\dot{M}_{out}$ is the mass loss rate; $l_{out}$ and $v_{out}$ are the spatial scale and the velocity of the observed outflow; $n_p$ and $n_e$ are the proton and electron number densities, respectively, in units of cm$^{-3}$; $V$ is the volume of the H$\alpha$ emission region. Equation (B2) is derived adopting the recombination rate in Osterbrock (1989) assuming an electron temperature of 10$^4$ K (see Appendix A). We also assume that collisional excitation of H$\alpha$ is negligible, since its transition level is too high to be thermally excited at the assumed temperature (Osterbrock 1989).
Using equations (B1) and (B2), the mean electron density is derived as follows:
\begin{eqnarray}
\langle n_e\rangle &\equiv& \frac{\int n_p n_e~dV}{\int n_p ~dV} = \frac{L_{H\alpha}~\rm{[erg~s^{-1}]}}{3.5 \times 10^{-25}} \frac{m_p v_{out}}{\dot{M} _{out} l_{out} \langle\chi_i\rangle}  \nonumber \\
&\sim & 1.9\times 10^{10} ~\langle\chi_i\rangle ^{-1} \left( \frac{L_{H\alpha}}{\rm 1~L_\odot} \right) \left( \frac{v_{out}}{\rm 100~km~s^{-1}} \right) \nonumber \\
& & ~~~~~ \left( \frac{\dot{M}_{out}}{\rm 10^{-8}~M_\odot~yr^{-1}} \right)^{-1} \left( \frac{l_{out}}{\rm 1~AU} \right)^{-1} ~\rm{cm^{-3}}
\end{eqnarray}
where the mean ionization fraction is defined as $\langle\chi_i\rangle \equiv (\int n_p ~dV)/(\int n_H ~dV)$.
From the equation (B3), the mean hydrogen number density is derived as follows:
\begin{eqnarray}
\langle n_H\rangle &\equiv&  \langle n_e\rangle / \langle \chi_i\rangle \nonumber \\
&\sim&  1.9\times 10^{10} ~ \langle \chi_i\rangle ^{-2} \left( \frac{L_{H\alpha}}{\rm 1~L_\odot} \right) \left( \frac{v_{out}}{\rm 100~km~s^{-1}} \right) \nonumber \\
& & ~~~~ \left( \frac{\dot{M}_{out}}{\rm 10^{-8}~M_\odot~yr^{-1}} \right)^{-1} \left( \frac{l_{out}}{\rm 1~AU} \right)^{-1} ~\rm{cm^{-3}}
\end{eqnarray}
Next, we derive the typical opening angle of the flow assuming axisymmetric geometry and that the flow has zero width at its origin. The total mass in the flow is described as follows:
\begin{eqnarray}
M_{out} &\sim& \int \pi r_{out}^2 m_p n_H f_H ~dl \nonumber \\
&\sim& 2 \pi \langle r_{out}^2 \rangle  l_{out} m_p \langle n_H \rangle \langle f_H \rangle
\end{eqnarray}
where $r_{out}$ is the radius (or half-width) of the flow at a certain distance from the star; $\langle r_{out} ^2 \rangle$ is the mean square of the radius; $f_H$ is the volume filling factor. On the other hand, the typical flow width $\langle d_{out} \rangle$ and opening angle $\langle \theta \rangle$ can be described as follows:
\begin{eqnarray}
\langle d_{out} \rangle &\equiv& 2 \sqrt{\langle r_{out}^2 \rangle} \\
\langle \theta \rangle &\equiv& 2 ~\rm{tan^{-1}} \frac{\it{\langle d_{out} \rangle}}{2 \it{l_{out}}},
\end{eqnarray}
Substituting equations (B1)(B5)(B6) to (B7), $\langle \theta \rangle$ is derived as follows:
\begin{eqnarray}
\langle \theta \rangle &\sim& 2 ~\rm{tan^{-1}} \it{\sqrt{ \frac{\dot{M}_{out}}{\rm{2} \it{\pi} m_p \langle n_H\rangle v_{out} l_{out}^{\rm{2}} \langle f_H \rangle}}},
\end{eqnarray}

\end{document}